\documentclass[11pt,a4paper]{article}
\pdfoutput=1
\usepackage{jheppub}

\usepackage{multirow, graphicx,amssymb,url,mathrsfs,amsmath}
\usepackage{wrapfig,boxedminipage,setspace,subfigure,epsfig}
\usepackage{amsxtra,amstext,latexsym,dsfont,amsfonts}
\usepackage{color,eucal}
\usepackage[dvipsnames]{xcolor}
\usepackage{float}
\usepackage{slashed,comment}
\usepackage{kotex}
\usepackage{tikz}
\usetikzlibrary{calc,patterns,angles,quotes}
\usetikzlibrary{decorations.pathreplacing,decorations.markings,snakes}
\usepackage{tabularx, array}
\usepackage{mdframed,mathtools}
\newcolumntype{L}[1]{>{\raggedright\arraybackslash}p{#1}}
\newcolumntype{C}[1]{>{\centering\arraybackslash}p{#1}}
\newcolumntype{R}[1]{>{\raggedleft\arraybackslash}p{#1}}

\newcommand{\ket}[1]{\mbox{$| #1 \rangle$}}

\newcommand{\be}{\begin{equation}}
\newcommand{\ee}{\end{equation}}
\newcommand{\bea}{\begin{eqnarray}}
\newcommand{\eea}{\end{eqnarray}}

\title{Lanczos Meets Orthogonal Polynomials}

\author[a,b]{Le-Chen Qu}

\emailAdd{lechen.qu@ift.csic.es}

\preprint{IFT-UAM/CSIC-25-161}

\affiliation[a]{Instituto de F\'isica Te\'orica UAM/CSIC, Calle Nicol\'as Cabrera 13-15, 28049 Madrid, Spain}
\affiliation[b]{Departamento de F\'isica Te\'orica, Universidad Aut{\'o}noma de Madrid, 28049 Madrid, Spain}

\abstract{
We establish a direct correspondence between the Lanczos approach and the orthogonal polynomials approach in random matrix theory. In the large-$N$ and continuum limits, the average Lanczos coefficients and the recursion coefficients become equivalent, with the precise mapping $b(1-x)=\sqrt{R(x)}$ and $a(1-x)=S(x)$. As a result, the two formalisms yield identical expressions for the leading density of states. We further analyze the Krylov dynamics associated with the recursion coefficients and show that the orthogonal polynomials admit a natural interpretation as Krylov polynomials. This picture is realized explicitly in the Gaussian Unitary Ensemble, where all quantities can be computed analytically.}

\begin{document}
\maketitle

%

\section{Introduction}
The study of quantum dynamics and random matrix theory (RMT) has uncovered profound connections between operator growth and spreading, and the statistical structure of quantum chaos. Among the analytical tools developed to probe these relationships, the \textit{Lanczos approach} plays a distinguished and central role. Originally introduced as a numerical algorithm for tridiagonalizing Hermitian operators \cite{Parker:2018yvk}, the Lanczos approach has evolved into a powerful theoretical framework for exploring Hilbert-space evolution. In particular, it provides a natural framework for describing the spreading of states under Hamiltonian dynamics in the Krylov basis, with the expansion coefficients encoding the associated correlation functions and giving rise to the notion of \textit{Krylov complexity} \cite{Barbon:2019wsy,Avdoshkin:2019trj,Rabinovici:2020ryf,Jian:2020qpp,Dymarsky:2021bjq,Hornedal:2022pkc,Balasubramanian:2022tpr,Caputa:2023vyr,Erdmenger:2023wjg,Craps:2023ivc,He:2024hkw,He:2024xjp,Caputa:2024vrn,Baggioli:2024wbz,Craps:2024suj,Huh:2024ytz,Caputa:2024sux,Zhai:2024tkz,Nandy:2024mml,Li:2024ljz,Balasubramanian:2024ghv,Bhattacharya:2024szw,Bhattacharya:2024hto,Bhattacharya:2024uxx,Aguilar-Gutierrez:2024nau,Craps:2025kub,Evnin:2025cfx,He:2025guu,Zhai:2025abc,Caputa:2025dep,Caputa:2025ozd,Caputa:2025mii,Miyaji:2025yvm,Takahashi:2025iol,Miyaji:2025ucp,Demulder:2025uda}. Building on this perspective, Krylov complexity has been argued to be closely related to circuit complexity \cite{Nielsen:2005mkt,Nielsen:2006cea,Dowling:2006tnk,Jefferson:2017sdb,Khan:2018rzm,Hackl:2018ptj,Chapman:2018hou,Bhattacharyya:2018bbv,Guo:2018kzl,Bernamonti:2019zyy,Ali:2019zcj,Caceres:2019pgf,Bernamonti:2020bcf,Ruan:2020vze,Qu:2021ius,Qu:2022zwq,Haque:2024ldr} as well as to holographic complexity \cite{Susskind:2014rva,Stanford:2014jda,Brown:2015bva,Brown:2015lvg,Cai:2016xho,Belin:2021bga,Belin:2022xmt,Carrasco:2023fcj,Jorstad:2023kmq,Jiang:2023jti,Caceres:2023ziv,Myers:2024vve,Arean:2024pzo,Jiang:2025qai,Miyaji:2025jxy,Caceres:2025myu,Caceres:2025ypk,Fatemiabhari:2025cyy,Fatemiabhari:2025usn,Fatemiabhari:2025poq}. For comprehensive reviews, see~\cite{Nandy:2024htc,Baiguera:2025dkc,Rabinovici:2025otw}.

Within the context of RMT, the Lanczos approach has revealed an unexpected bridge between spectral statistics and unitary dynamics. As shown in Ref.~\cite{Balasubramanian:2022dnj}, the ensemble-averaged tridiagonal elements, known as the average \textit{Lanczos coefficients}, are directly related to the leading spectral density through Eq.~\eqref{densityofstatesandtheLanczoscoefficients}. This result establishes a precise correspondence between the eigenvalue distribution and the statistical properties of the Lanczos coefficients, thereby linking spectral information to the dynamics of quantum state spreading in Hilbert space. Recent progress in understanding these connections has been reported in Ref.~\cite{Balasubramanian:2023kwd,Balasubramanian:2024lqk,Nandy:2024zcd,Loc:2024oen}.

{In parallel, the \textit{orthogonal polynomials approach} has long been fundamental to the analytical study of RMT \cite{BESSIS1980109,Ginsparg:1991bi,Ginsparg:1993is,bleher2011lectures,Eynard:2015aea, livan2018introduction}.} The recursion relations defining these polynomials encode the spectral information of the matrix ensembles in a compact algebraic form through the associated \textit{recursion coefficients}. The joint probability distribution of $m$ eigenvalues is given by the Dyson determinantal formula, Eq.~\eqref{Dysondeterminantalormula}. Integrating out all but one eigenvalue, the leading density of states can be expressed in terms of the recursion coefficients, as shown in Eq.~\eqref{leadingspectraldensity} and first established in Ref.~\cite{Dalley:1990hb}. The similarity between Eqs.~\eqref{densityofstatesandtheLanczoscoefficients} and~\eqref{leadingspectraldensity} naturally raises the question of whether there exists a direct correspondence between the average Lanczos coefficients in the Lanczos approach and the recursion coefficients in the orthogonal polynomials approach.

As we show below, such a correspondence indeed exists. In the large-$N$ and continuum limits, we demonstrate that the average Lanczos coefficients are directly related to the recursion coefficients through Eq.~\eqref{mainresu1}. For finite $N$, the two sets of equations differ only by a factor in the numerator, namely \(2n\) versus \(2(N-n)-1\).  We then analyze the Krylov dynamics associated with the recursion coefficients\footnote{It is worth emphasizing that this dynamics differs from that associated with the average Lanczos coefficients.}. The orthogonal polynomials admit a natural interpretation as \emph{Krylov polynomials}, allowing techniques from both the orthogonal polynomials approach and the Lanczos approach to be used in a unified way. As a concrete illustration, we conclude by examining the Gaussian Unitary Ensemble (GUE), where all results can be derived analytically and the correspondence between the two approaches is fully exhibited.

This paper is structured as follows. 
Section~\ref{sec2} provides essential background on the Lanczos approach within RMT. In Section~\ref{sec3}, we review the framework of the orthogonal polynomials approach and demonstrate a direct correspondence between the average Lanczos coefficients and the recursion coefficients. Section~\ref{sec4} discusses the Krylov dynamics associated with the recursion coefficients and examines the GUE as a concrete example. Finally, Section~\ref{sec5} presents our conclusions and outlines possible directions for future research.

\section{The Lanczos approach}\label{sec2}

In this section, the framework of the Lanczos approach is reviewed within the context of RMT \cite{Balasubramanian:2022dnj,Balasubramanian:2022tpr}. Readers familiar with the material can skip this section.

\subsection{The Lanczos approach to quantum dynamics}

Considering a quantum system with a time-independent Hamiltonian $H$, the time evolution of a state $\ket{\psi(t)}$ is governed by
\begin{equation}
\label{Schr}
|\psi(t)\rangle = e^{-iHt}\,|\psi(0)\rangle\,,
\end{equation}
which expands as a linear combination of the states
\(
\{\,|\psi(0)\rangle,\; H|\psi(0)\rangle,\; H^2|\psi(0)\rangle,\; \ldots\,\}\,.
\)
Applying the Lanczos approach amounts to orthonormalizing this sequence via the Gram-Schmidt procedure with the standard inner product. The orthonormal vectors obtained in this way form the Krylov basis. In this basis the Hamiltonian takes a tridiagonal form,
\begin{equation}\label{tridiagham}
H =
\begin{pmatrix}
a_0     & b_1     & 0       & 0        & 0         & \cdots & 0 \\
b_1     & a_1     & b_2     & 0        & 0         & \cdots & 0 \\
0       & b_2     & a_2     & b_3      & 0         & \cdots & 0 \\
0       & 0       & b_3     & a_3      & b_4       & \cdots & 0 \\
\vdots  & \vdots  & \vdots  & \ddots   & \ddots    & \ddots & \vdots \\
0       & 0       & 0       & \cdots   & b_{N-2}   & a_{N-2} & b_{N-1} \\
0       & 0       & 0       & \cdots   & 0         & b_{N-1} & a_{N-1}
\end{pmatrix},
\end{equation}
and the sequences $\{a_n\}$ and $\{b_n\}$ are known as \textit{Lanczos coefficients} \cite{Balasubramanian:2022tpr}.

\subsection{Random matrix theory}
A random matrix theory is defined by an ensemble of $N\times N$ Hermitian matrices $H_{ij}$ with a probability measure determined by a potential $V(H)$ \cite{Eynard:2015aea, bleher2011lectures, livan2018introduction},
\begin{equation}
    d\mu_N(H) = \frac{1}{Z_N}\, e^{-N\,\mathrm{Tr}\,V(H)}\, dH ,
\end{equation}
where the potential $V(\lambda)$ is assumed to be a polynomial satisfying
\begin{equation}\label{boundarycondition}
    \lim_{\lambda\to\pm\infty} \bigl( V(\lambda) - \log(\lambda^2+1) \bigr) = +\infty ,
\end{equation}
so that the measure is normalizable. The Lebesgue measure on the space of Hermitian matrices is
\begin{equation}
    dH = \prod_{j=1}^N dH_{jj} \prod_{j<k} d\,\mathrm{Re}\,H_{jk}\, d\,\mathrm{Im}\,H_{jk},
\end{equation}
and the normalization factor
\begin{equation}
    Z_N = \int e^{-N\,\mathrm{Tr}\,V(H)}\, dH
\end{equation}
is the partition function of the ensemble. The measure $d\mu_N(H)$ is invariant under unitary conjugation,
\begin{equation}
    H \;\longrightarrow\; U^{-1}HU , \qquad U\in U(N),
\end{equation}
which characterizes the ensemble. By the Weyl integral formula, this invariance implies that the eigenvalues of $H$ are distributed according to the joint probability density
\begin{equation}
    d\mu_N(\lambda)
    = \frac{1}{\tilde{Z}_N}\,
      \prod_{j>k}(\lambda_j - \lambda_k)^2
      \exp\!\left[-N\sum_{i=1}^N V(\lambda_i)\right]
      \prod_{i=1}^N d\lambda_i ,
\end{equation}
where the normalization constant is
\begin{equation}
    \tilde{Z}_N
    = \int
      \prod_{j>k}(\lambda_j - \lambda_k)^2
      \exp\!\left[-N\sum_{i=1}^N V(\lambda_i)\right]
      \prod_{i=1}^N d\lambda_i .
\end{equation}

\subsection{The Lanczos approach to random matrix theory}
Once a Hamiltonian $H$ and an initial state $|\psi\rangle$ are specified, the Lanczos procedure produces the tridiagonal representation shown in~\eqref{tridiagham}. In the RMT, however, one begins with an ensemble of Hamiltonians, so the Lanczos algorithm naturally generates an ensemble of tridiagonal random matrices. The joint probability distribution is given by \cite{Balasubramanian:2022dnj}
\begin{equation}
p(a_0,\ldots,a_{N-1},\, b_1,\ldots,b_{N-1})
\propto 
\left(\prod_{n=1}^{N-1} b_n^{\,2(N-n)-1}\right)
\exp\!\left[-N\,\mathrm{Tr}\,V(H)\right].
\end{equation}
In the large-$N$ limit, the average Lanczos coefficients follow from a saddle-point analysis of the effective action
\begin{equation}\label{SLan}
    S_{\mathrm{Lanczos}}
    = \sum_{n}\!\left[(2(N-n)-1)\ln b_{n}\right]
      - N\,\mathrm{Tr}\,V(H)\, .
\end{equation}
Introducing the continuum variable $x=n/N$ and treating $\{a_{n},b_{n}\}$ as smooth functions $\{a(x),b(x)\}$, the saddle-point equations that extremize this action become
\begin{equation}\label{extremaofthepotential}
\begin{aligned}
    2(1 - x) 
    &= b(x)\,\frac{\partial}{\partial b(x)}
    \left[
        \int d\lambda\,
        \frac{V(\lambda)}{\pi\sqrt{4 b(x)^{2} - (\lambda - a(x))^{2}}}
    \right],
    \\[4pt]
    0 
    &= \frac{\partial}{\partial a(x)}
    \left[
        \int d\lambda\,
        \frac{V(\lambda)}{\pi\sqrt{4 b(x)^{2} - (\lambda - a(x))^{2}}}
    \right].
\end{aligned}
\end{equation}
For a polynomial potential
\(
    V(\lambda)=\sum_{m} w_m \lambda^{m},
\)
the integrals that enter \eqref{extremaofthepotential} can be evaluated analytically. The saddle-point equations then reduce to algebraic relations between $a(x)$ and $b(x)$\footnote{A typo in Eq.~(89) of Ref.~\cite{Balasubramanian:2022dnj} was later corrected in Eq.~(5.8) of Ref.~\cite{Nandy:2024zcd}.},
\begin{equation}\label{polyabv}
\begin{aligned}
    0
    &= \sum_{m} w_m
       \sum_{q=0}^{m-1}
       (m-q)\, a^{\,m-q-1} b^{\,q}
       \binom{m}{q}\binom{q}{q/2},
    \\
    1 - x 
    &= \frac{1}{2}
       \sum_{m} w_m 
       \sum_{q=2}^{m-1}
       q\, a^{\,m-q} b^{\,q}
       \binom{m}{q}\binom{q}{q/2}.
\end{aligned}
\end{equation}
Once these equations are solved, the leading density of states follows directly from the resulting functions \(a(x)\) and \(b(x)\), see Ref.~\cite{Balasubramanian:2022dnj}, 
\begin{equation}\label{densityofstatesandtheLanczoscoefficients}
    \rho_0(\lambda)
    =
    \frac{1}{\pi}
    \int_{0}^{1} dx\,
    \frac{
        \Theta\!\left(4 b(x)^{2} - (\lambda - a(x))^{2}\right)
    }{
        \sqrt{4 b(x)^{2} - (\lambda - a(x))^{2}}
    },
\end{equation}
where $\Theta(x)$ is the Heaviside step function.

\section{The orthogonal polynomials approach}\label{sec3}
In this section, we review the orthogonal polynomials approach to random matrix theory \cite{BESSIS1980109,Ginsparg:1991bi,Eynard:2015aea, bleher2011lectures,Ginsparg:1993is,livan2018introduction} and demonstrate a direct correspondence between the average Lanczos coefficients and the recursion coefficients.

\subsection{Orthogonal polynomials and joint eigenvalue distributions}
A complementary, and often more powerful, approach to RMT is built on a family of monic \textit{orthogonal polynomials}
\(
P_n(\lambda) = \lambda^{n} + \text{(lower-order terms)},
\)
defined with respect to the inner product
\begin{equation}\label{innerproduct}
    \int_{-\infty}^{\infty}
    P_n(\lambda)\, P_m(\lambda)\,
    e^{-N V(\lambda)}\, d\lambda
    = h_n\, \delta_{mn}\, .
\end{equation}
The constants \(h_n\) follow from the requirement that each \(P_n\) has unit leading coefficient. In particular,
\(
h_0 = \int_{-\infty}^{\infty} d\mu(\lambda),
\)
where the measure
\(
d\mu(\lambda) = e^{-N V(\lambda)}\, d\lambda
\)
is determined by the potential \(V(\lambda)\). A useful representation of the orthogonal polynomials is given by \emph{Heine's formula}, which expresses the $n$-th monic polynomial as the ensemble average of a characteristic determinant:
\begin{equation}\label{Heineformula}
P_n(\lambda)
= \mathbb{E}\!\left[\det(\lambda I_n - H)\right]
=
\frac{
\int
\prod_{i=1}^{n} (\lambda - \lambda_i)\,
\prod_{i<j} |\lambda_i - \lambda_j|^{2}\,
e^{-N \sum_{i=1}^{n} V(\lambda_i)}\,
d\lambda_1 \cdots d\lambda_n
}{
\int
\prod_{i=1}^{n} d\lambda_i\,
\prod_{i<j} |\lambda_i - \lambda_j|^{2}\,
e^{-N \sum_{i=1}^{n} V(\lambda_i)}
}\, .
\end{equation}
Here $\mathbb{E}[\cdots]$ denotes the expectation with respect to the joint eigenvalue distribution of $H$, and $I_n$ is the $n\times n$ identity matrix. Recall that arbitrary polynomials can be generated by adding linear combinations of earlier columns in a determinant, a transformation that leaves the determinant unchanged. As a result, the Vandermonde determinant may be written as
\begin{equation}
\Delta(\lambda)
= \prod_{i<j} (\lambda_j - \lambda_i)
= \det(\lambda_i^{\,j-1})
= \det\!\left[P_{j-1}(\lambda_i)\right],
\end{equation}
where $P_n(\lambda)$ denotes the $n$-th orthogonal polynomial. Using this identity, the joint distribution of $m$ eigenvalues can be expressed entirely in terms of the first $N$ orthogonal polynomials through the \textit{Dyson determinantal formula},
\begin{equation}\label{Dysondeterminantalormula}
\begin{aligned}
R_m(\lambda_1,\ldots,\lambda_m)
&= \frac{N!}{\tilde{Z}_N (N-m)!}
\int \prod_{j>k} (\lambda_j - \lambda_k)^2\,
e^{-N \sum_{i=1}^N V(\lambda_i)}
\, d\lambda_{m+1}\cdots d\lambda_N
\\[4pt]
&=
\det\!\left[
\sum_{n=0}^{N-1} \frac{1}{h_n}\,
P_n(\lambda_k)\, P_n(\lambda_l)\,
e^{-\frac{N}{2} V(\lambda_k)}
e^{-\frac{N}{2} V(\lambda_l)}
\right]_{k,l = 1}^{m}.
\end{aligned}
\end{equation}
In particular, the density of states, or one-point function, follows as
\begin{equation}\label{spectraldensity}
\rho(\lambda)
= \frac{1}{N} R_1(\lambda)
= e^{-N V(\lambda)}\, \frac{1}{N}
\sum_{n=0}^{N-1} \frac{1}{h_n}\, P_n^{2}(\lambda)\, .
\end{equation}

\subsection{Recurrence relation and discrete string equations} 
The monic orthogonal polynomials $P_n(\lambda)$ satisfy the three-term recursion relation
\begin{equation}\label{recursionrelation}
    \lambda\, P_n(\lambda)
    = P_{n+1}(\lambda)
      + S_n\, P_n(\lambda)
      + R_n\, P_{n-1}(\lambda)\, ,
\end{equation}
where $S_n$ and $R_n$ are the \textit{recursion coefficients}. This truncation to three terms follows directly from the orthogonality condition~\eqref{innerproduct}, which ensures that $\lambda P_n$ has components only along $P_{n+1}$, $P_n$, and $P_{n-1}$. It is convenient to introduce an auxiliary Hilbert space spanned by the normalized states
\begin{equation}
    |n\rangle = \frac{P_n(\lambda)}{\sqrt{h_n}}, \qquad n = 0,1,2,\ldots ,
\end{equation}
which amounts to rescaling the inner product in~\eqref{innerproduct} so that the states form an orthonormal basis:
\begin{equation}
    \langle n | m \rangle = \delta_{mn}.
\end{equation}
Within the auxiliary Hilbert space, the recursion relation~\eqref{recursionrelation} is given by
\begin{equation}\label{compactrecursionrelation}
    \hat{\lambda}\, |n\rangle
    = \sqrt{R_{n+1}}\, |n+1\rangle
      + S_n\, |n\rangle
      + \sqrt{R_n}\, |n-1\rangle,
\end{equation}
so that the operator $\hat{\lambda}$ is represented in this basis by the real symmetric tridiagonal Jacobi matrix
\begin{equation}\label{tridiagonalJacobilam}
\hat{\lambda} =
\begin{pmatrix}
S_0 & \sqrt{R_1} & 0 & 0 & \cdots \\
\sqrt{R_1} & S_1 & \sqrt{R_2} & 0 & \cdots \\
0 & \sqrt{R_2} & S_2 & \sqrt{R_3} & \cdots \\
0 & 0 & \sqrt{R_3} & S_3 & \cdots \\
\vdots & \vdots & \vdots & \vdots & \ddots
\end{pmatrix}.
\end{equation}
Relations among the recursion coefficients follow from the two identities
\begin{equation}
\begin{aligned}
    \int d\mu(\lambda)\, P_n(\lambda)\, \frac{d}{d\lambda} P_n(\lambda) &= 0,\\
    \int d\mu(\lambda)\, P_{n-1}(\lambda)\, \frac{d}{d\lambda} P_n(\lambda) &= n\, h_{n-1}.
\end{aligned}
\end{equation}
Integrating by parts and using the vanishing of boundary terms due to~\eqref{boundarycondition}, the derivative acting on the exponential part of the measure produces an insertion of $-N V'(\lambda)$. In this way the integral identities are rewritten as constraints on matrix elements of \(V'(\lambda)\) in the auxiliary Hilbert space, yielding the \emph{discrete string equations} for the recursion coefficients,
\begin{equation}\label{VSRREEQ}
\begin{aligned}
\langle n | V'(\hat{\lambda}) | n \rangle &= 0 ,\\
\sqrt{R_n}\,\langle n-1 | V'(\hat{\lambda}) | n \rangle &= \frac{n}{N} .
\end{aligned}
\end{equation}

\subsection{Leading density of states via orthogonal polynomials} 
The moments of the density of states can be written in terms of the orthonormal states \(|n\rangle\) as \cite{Dalley:1990hb,Dalley:1990zs,Dalley:1991jp}
\begin{equation}
\int d\lambda\, \rho(\lambda)\, \lambda^{p}
= \frac{1}{N} \sum_{n=0}^{N-1}
\int d\mu(\lambda)\, \frac{1}{h_n}\, P_n^{2}(\lambda)\, \lambda^{p}
= \frac{1}{N} \sum_{n=0}^{N-1} \langle n | \hat{\lambda}^{\,p} | n \rangle .
\end{equation}
To evaluate these matrix elements, we introduce raising and lowering operators acting on the orthonormal basis,
\(
\hat{L}_{+}|n\rangle = |n+1\rangle
\)
and
\(
\hat{L}_{-}|n\rangle = |n-1\rangle .
\)
Using the recursion relation~\eqref{compactrecursionrelation}, the operator \(\hat{\lambda}\) takes the form
\begin{equation}
\hat{\lambda}
= \sqrt{R_{n+1}}\, \hat{L}_{+}
+ \sqrt{R_{n}}\, \hat{L}_{-}
+ S_{n} .
\end{equation}
In the large-\(N\) limit we expand \(\hat{\lambda}^{p}\) while neglecting operator ordering, then the dominant contributions arise from terms with equal numbers of raising and lowering operators. This is counted by binomial coefficients
\begin{equation}
\begin{aligned}
\int d\lambda\, \rho(\lambda)\, \lambda^{p}
&= \frac{1}{N} \sum_{n=0}^{N-1}
\langle n |
\sum_{r=0}^{p} \binom{p}{r}\,
S_{n}^{\,p-r}\,
\left( \sqrt{R_{n+1}}\, \hat{L}_{+}
     + \sqrt{R_{n}}  \,\hat{L}_{-} \right)^{r}
| n \rangle
\\
&\approx
\frac{1}{N} \sum_{n=0}^{N-1}
\sum_{r=0}^{p} \binom{p}{r}\binom{r}{r/2}\,
S_{n}^{\,p-r}\, R_{n}^{\,r/2}
\\
&\approx
\frac{1}{N} \sum_{n=0}^{N-1}
\int_{-2}^{2}
\frac{dy}{\pi \sqrt{4-y^{2}}}\,
\bigl( y\sqrt{R_{n}} + S_{n} \bigr)^{p},
\end{aligned}
\end{equation}
where we used the identity
\(
\binom{m}{m/2}
= \int_{-2}^{2}\frac{y^{m}}{\pi\sqrt{4-y^{2}}}\,dy
\),
and approximated \(R_{n+1}\approx R_{n}\), \(S_{n+1}\approx S_{n}\) at large \(N\). In the continuum limit with \(x = n/N\), the leading density of states follows as \cite{Dalley:1990hb},
\begin{equation}\label{leadingspectraldensity}
\begin{aligned}
    \rho_{0}(\lambda)
=& \frac{1}{\pi}
\int_{0}^{1} dx
\int_{-2}^{2}
\frac{dy}{\sqrt{4 - y^{2}}}\,
\delta\!\left(\lambda - \bigl[ y\sqrt{R(x)} + S(x) \bigr] \right)\\
=& \frac{1}{\pi} \int_{0}^{1} dx\,
\frac{
\Theta\!\left[ 4R(x) - (\lambda - S(x))^{2} \right]
}{
\sqrt{\,4R(x) - (\lambda - S(x))^{2}\,}
},
\end{aligned}
\end{equation}
which expresses the leading density of states directly in terms of the recursion coefficients $\{R(x),S(x)\}$. It is noteworthy that equations~\eqref{densityofstatesandtheLanczoscoefficients} and~\eqref{leadingspectraldensity} share the same structural form. This similarity suggests that the two descriptions may not be independent. In particular, it raises the question of whether the average Lanczos coefficients $\{a(x), b(x)\}$ can be directly identified with the recursion coefficients $\{R(x), S(x)\}$ that appear in the orthogonal polynomials approach.

\subsection{Equivalence between average Lanczos coefficients and recursion coefficients} 
In order to compare the recursion coefficients with the average Lanczos coefficients, we begin by simplifying the discrete string equations. The first discrete string equation~\eqref{VSRREEQ} may be simplified in the same manner as the moment expansion discussed earlier.  
For a polynomial potential, $V(\hat{\lambda}) = \sum_m w_m \hat{\lambda}^{m}$, we use the expansion of \(\hat{\lambda}\) in terms of \(S_n\) and \(R_n\) to obtain
\begin{equation}\label{firstdiscretestringequation}
\begin{aligned}
0
&= \langle n | V'(\hat{\lambda}) | n \rangle
= \sum_m m w_m \langle n | \hat{\lambda}^{m-1} | n \rangle \\
&= \sum_m m w_m 
\langle n |
\sum_{q=0}^{m-1} 
\binom{m-1}{q}\,
S_n^{\,m-q-1}
\left( \sqrt{R_{n+1}}\,\hat{L}_{+}
     + \sqrt{R_n}\,\hat{L}_{-} \right)^{q}
| n \rangle \\
&\approx \sum_m m w_m 
\sum_{q=0}^{m-1}
\binom{m-1}{q}\binom{q}{q/2}\,
S_n^{\,m-q-1} R_n^{\,q/2} \\
&\approx \sum_m w_m 
\sum_{q=0}^{m-1} (m - q)\,
S^{m-q-1} R^{q/2}
\binom{m}{q}\binom{q}{q/2},
\end{aligned}
\end{equation}
where we used
\[
\binom{m-1}{q}\binom{q}{q/2}
= \frac{m-q}{m}\binom{m}{q}\binom{q}{q/2},
\]
and noted that only even \(q\) contribute. The second discrete string equation involves off-diagonal elements of \(V'(\hat{\lambda})\).  
These correspond combinatorially to paths that start at \(|n\rangle\), end at \(|n-1\rangle\), and contain \(q/2\) lowering operations, \(q/2-1\) raising operations, and \(m-q\) stationary steps.  
This yields
\begin{equation}\label{seconddiscretestringequation}
\begin{aligned}
x &= \frac{n}{N}
= \sqrt{R_n}\langle n-1 | V'(\hat{\lambda}) | n \rangle
= \sum_m m w_m \sqrt{R_n}\langle n-1 | \hat{\lambda}^{m-1} | n \rangle \\
&= \sum_m m w_m 
\langle n-1 |
\sum_{q=2}^{m-1}
\binom{m-1}{q-1}\,
\sqrt{R_n}\, S_n^{\,m-q}
\left( \sqrt{R_{n+1}}\,\hat{L}_{+}
     + \sqrt{R_n}\,\hat{L}_{-} \right)^{q-1}
| n \rangle \\
&\approx
\sum_m m w_m
\sum_{q=2}^{m-1}
\binom{m-1}{q-1} \binom{q-1}{q/2}\,
S_n^{\,m-q} R_n^{\,q/2} \\
&\approx 
\frac{1}{2}\sum_m w_m
\sum_{q=2}^{m-1}
q\, S^{m-q} R^{q/2}
\binom{m}{q}\binom{q}{q/2},
\end{aligned}
\end{equation}
where 
\[
\binom{m-1}{q-1}\binom{q-1}{q/2}
= \frac{q}{2m}\binom{m}{q}\binom{q}{q/2},
\]
and again only even \(q\) contribute. Comparing equations~\eqref{firstdiscretestringequation} and~\eqref{seconddiscretestringequation} with the saddle-point equations governing the average Lanczos coefficients in~\eqref{polyabv}, we obtain the identifications
\begin{equation}\label{mainresu1}
\sqrt{R(x)} = b(1 - x), 
\qquad
S(x) = a(1 - x).
\end{equation}
Thus the recursion coefficients $\{\sqrt{R(x)},S(x)\}$ map directly onto the average Lanczos coefficients $\{b(x),a(x)\}$, modulo a reversal of the continuum coordinate \(x \to 1-x\). This is fully consistent with the observation that the leading densities of states in~\eqref{densityofstatesandtheLanczoscoefficients} and~\eqref{leadingspectraldensity} exhibit the same structural form, as reversing the integration variable does not alter the resulting expression. Equation \eqref{mainresu1} is one of the main results of this article. A natural question is whether the recursion coefficients \(\{R_n, S_n\}\) and the average Lanczos coefficients \(\{b_n, a_n\}\) satisfy equations that share the same structural form even before taking the continuum limit. To address this, we introduce the effective action for recursion coefficients 
\(\{R_n, S_n\}\), following Ref.~\cite{bleher2008lecturesrandommatrixmodels}, 
\begin{equation}
S_{\mathrm{Polynomial}}
= \sum_{n=1}^{N} n \ln R_n 
  - N\, \mathrm{Tr}\, V(\hat{\lambda}) .
\end{equation}
Extremizing this action yields
\begin{equation}\label{extreme1}
\begin{aligned}
\frac{\partial S_{\mathrm{Polynomial}}}{\partial \sqrt{R_n}}
&= \frac{2n}{\sqrt{R_n}}
   - N\, \mathrm{Tr}\!\left[
       V'(\hat{\lambda}) 
       \frac{\partial \hat{\lambda}}{\partial \sqrt{R_n}}
     \right]
 = \frac{2n}{\sqrt{R_n}}
   - 2N\, \langle n-1 | V'(\hat{\lambda}) | n \rangle = 0 , 
\\[4pt]
\frac{\partial S_{\mathrm{Polynomial}}}{\partial S_n}
&= -N\, \mathrm{Tr}\!\left[
      V'(\hat{\lambda}) 
      \frac{\partial \hat{\lambda}}{\partial S_n}
    \right]
 = -N\, \langle n | V'(\hat{\lambda}) | n \rangle = 0 .
\end{aligned}
\end{equation}
These equations reproduce exactly the discrete string equations~\eqref{VSRREEQ}. An analogous extremization may be applied to the Lanczos effective action \(S_{\mathrm{Lanczos}}\) defined in~\eqref{SLan}.  
Varying with respect to \(b_n\) and \(a_n\), and using the dependence of the tridiagonal matrix \(H\) on these coefficients \eqref{tridiagham}, we obtain
\begin{equation}\label{extreme2}
\begin{aligned}
\frac{\partial S_{\mathrm{Lanczos}}}{\partial b_n}
&= \frac{2(N-n)-1}{b_n}
   - N\, \mathrm{Tr}\!\left[
       V'(H) \frac{\partial H}{\partial b_n}
     \right]
 = \frac{2(N-n)-1}{b_n}
   - 2N\, V'(H)_{n-1,n} = 0 ,
\\[4pt]
\frac{\partial S_{\mathrm{Lanczos}}}{\partial a_n}
&= -N\, \mathrm{Tr}\!\left[
      V'(H) \frac{\partial H}{\partial a_n}
    \right]
 = -N\, V'(H)_{n,n} = 0 .
\end{aligned}
\end{equation}
Comparing the extremization equations from the polynomial action~\eqref{extreme1} and the Lanczos action~\eqref{extreme2}, we note that both \(V'(\hat{\lambda})\) and \( V'(H)\) share the same structure. Consequently, the equations involving diagonal matrix elements coincide exactly in the two formulations.  
The equations involving off-diagonal elements differ only by a prefactor in the numerator, namely \(2n\) versus \(2(N-n)-1\)\footnote{{Note that the factor $2(N-n)-1$ arises from the saddle-point approximation employed in the evaluation of the average Lanczos coefficients. In the case of the Gaussian Unitary Ensemble, however, the average Lanczos coefficients can be computed exactly at finite $N$, without invoking the saddle-point approximation. In this case, the resulting coefficients are in exact agreement with the recursion coefficients, as shown in Eqs.~\eqref{bnensve} and \eqref{snrngue}. It would be interesting to investigate whether a similarly exact correspondence persists in more general matrix ensembles, which we leave for future work.}}. Crucially, this discrepancy vanishes in the continuum limit, where the recursion coefficients \(\{R(x), S(x)\}\) and the average Lanczos coefficients \(\{b(x), a(x)\}\) reproduce the correspondence of Eq.~\eqref{mainresu1}. This result suggests a mapping between two levels of description: the average Lanczos coefficients, which encode fine-grained dynamics and require ensemble averaging, and the recursion coefficients, which provide an intrinsic coarse-grained description independent of averaging. This correspondence closely parallels the emergence of the chord Hilbert space in the double-scaled Sachdev–Ye–Kitaev (DSSYK) model, where coarse-grained degrees of freedom have been suggested to capture an effective low-energy description of wormhole geometries~\cite{Berkooz:2018qkz,Lin:2022rbf,Rabinovici:2023yex,Heller:2024ldz,Xu:2024gfm,Ambrosini:2025hvo,Fu:2025kkh}. An important advantage of the orthogonal polynomials approach is that it applies to general matrix potentials $V(H)$, including those relevant to DSSYK~\cite{Balasubramanian:2024lqk,Nandy:2024zcd}. It would therefore be very interesting to further develop this framework and explore whether the auxiliary Hilbert space associated with the recursion coefficients admits a geometric or gravitational interpretation in the context of DSSYK.

\section{Orthogonal polynomials as Krylov polynomials}\label{sec4}
In this section, we examine the Krylov dynamics governed by the recursion coefficients \(\{R_n, S_n\}\) and illustrate the framework using the Gaussian Unitary Ensemble, which serves as an analytically solvable example.

\subsection{Krylov dynamics from orthogonal polynomials} 
As is well known, Lanczos coefficients naturally appear in the dynamical evolution of quantum states.  
Given the identification~\eqref{mainresu1}, it is natural to ask whether the recursion coefficients 
\(\{R_n, S_n\}\) admit a similar dynamical interpretation. As we show below, the answer is affirmative: the recursion relation~\eqref{compactrecursionrelation} implies that the time evolution generated by 
\(\hat{\lambda}\) takes the form
\begin{equation}\label{krylovdyna}
|\psi(t)\rangle
= e^{-i\hat{\lambda} t} |0\rangle
= \sum_{n=0}^{\infty} \frac{(-i t)^{n}}{n!}\, \hat{\lambda}^{\,n} |0\rangle
= \sum_{n=0}^{\infty} \psi_n(t)\, |n\rangle ,
\end{equation}
where the Krylov amplitudes satisfy
\begin{equation}\label{psieq}
i\,\partial_t \psi_n(t)
= \sqrt{R_{n+1}}\,\psi_{n+1}(t)
 + S_n\,\psi_n(t)
 + \sqrt{R_n}\,\psi_{n-1}(t),
\qquad 
\psi_n(0)=\delta_{n0}.
\end{equation}
Since the states \(|n\rangle = P_n(\lambda)/\sqrt{h_n}\) form an orthonormal basis with respect to the measure \(d\mu(\lambda) = e^{-N V(\lambda)} d\lambda\), the polynomials \(P_n(\lambda)\) therefore define a natural Krylov basis, or equivalently, a family of \emph{Krylov polynomials}, while the recursion coefficients \(\{R_n, S_n\}\) can be reinterpreted as the Lanczos coefficients in this framework. Recent developments on Krylov polynomials include~\cite{Muck:2022xfc, Kar:2021nbm, Muck:2024fpb, Adhikari:2025vdl, Alishahiha:2024vbf,Balasubramanian:2025xkj}. From this perspective, various dynamical quantities can be extracted directly from the measure $d\mu(\lambda)$. For instance, the survival amplitude is its Fourier transform,
\begin{equation}\label{survivalamplitude}
S(t)
= \langle 0| e^{-i\lambda t} |0\rangle
= \frac{1}{h_0} \int e^{-i\lambda t}\, d\mu(\lambda),
\end{equation}
while the moments are
\begin{equation}\label{moments}
M_k
= \langle 0 | \lambda^k | 0 \rangle
= \frac{1}{h_0} \int \lambda^k\, d\mu(\lambda).
\end{equation}
The recursion coefficients \(\{R_n, S_n\}\) may be reconstructed from these moments via the Markov-chain algorithm in Ref.~\cite{Balasubramanian:2022tpr}.  
For an even potential \(V(\lambda)\), all odd moments vanish, so \(S_n=0\), and the remaining moments determine the sequence \(\{R_n\}\).  
In particular, the Hankel matrix \(M_{ij}=M_{i+j}\) obeys
\begin{equation}\label{OCF1}
R_1^{\,n} R_2^{\,n-1} \cdots R_n
= \det\!\left(M_{i+j}\right)_{0\le i,j\le n}.
\end{equation}
Equivalently, one may use the recursive formulas
\begin{equation}\label{OCF2}
\begin{aligned}
M_{2\ell}^{(j)}
&= \frac{M_{2\ell}^{(j-1)}}{R_{j-1}}
 - \frac{M_{2\ell-2}^{(j-2)}}{R_{j-2}},
\qquad \ell = j,\dots, n ,\\[4pt]
M_{2\ell}^{(0)} &= M_{2\ell}, \qquad 
R_{-1}=R_0 := 1, \qquad
M_{2\ell}^{(-1)} = 0 , \\[4pt]
R_n &= M_{2n}^{(n)} .
\end{aligned}
\end{equation}
These recurrence formulas offer an alternative way to reconstruct the recursion coefficients, complementing the discrete string equations. Conversely, the discrete string equations~\eqref{VSRREEQ}, together with Heine’s formula~\eqref{Heineformula}, provide an independent route for constructing the Krylov polynomials and extracting Lanczos coefficients.  
From this perspective, the monic Krylov polynomials generated by a measure \(d\mu(\lambda)=U(\lambda)\, d\lambda\) can be mapped to the orthogonal polynomials of a random matrix model with potential
\begin{equation}
    V(\lambda) = -\frac{\log{U(\lambda)}}{N}.
\end{equation}
This mapping enables us to apply the methods in Sec.~\ref{sec3} to analyze the associated Krylov polynomials. Finally, a key dynamical quantity in Krylov space is the \emph{spread complexity},
\begin{equation}
C(t)
= \sum_n n\, |\psi_n(t)|^2
= \sum_n \frac{n}{h_0 h_n}
  \int d\mu(\lambda)\, d\mu(\lambda')\,
P_n(\lambda)\, P_n(\lambda')\, 
e^{-i(\lambda - \lambda')t},
\end{equation}
which characterizes the growth of the state's support in the Krylov basis and thus the effective spread of information or complexity over time.

\subsection{Gaussian unitary ensemble} 
We now apply the framework developed above to a concrete and analytically solvable example: the Gaussian Unitary Ensemble, corresponding to the quadratic potential \(V(\lambda)=\lambda^{2}/2\). {A discussion of non-Gaussian ensembles is deferred to appendix~\ref{app:nongaussian}.} We begin with the Lanczos approach of Sec.~\ref{sec2}. As shown in Refs.~\cite{dumitriu2002matrix, Balasubramanian:2022dnj}, the Lanczos coefficients in the GUE are distributed as
\begin{equation}
p(a_n)=\frac{1}{\sqrt{2N}}\,\mathcal{N}(0,2)(a_n),
\qquad
p(b_n)=\frac{1}{\sqrt{2N}}\,\chi_{2(N-n)}(b_n),
\end{equation}
where \(\mathcal{N}(r,s)\) denotes the Gaussian distribution with mean \(r\) and variance \(s\), and $\chi_r(x)$ is the chi-distribution given by
\begin{equation}
p_{\chi_{r}}(x)
= \frac{1}{2^{r/2 - 1}\, \Gamma(r/2)} \,
x^{\, r-1} e^{-x^{2}/2} .
\end{equation}
Averaging over the ensemble yields the average Lanczos coefficients \cite{dumitriu2002matrix, Balasubramanian:2022dnj}
\begin{equation}\label{bnensve}
\langle a_n \rangle = 0,
\qquad
\langle b_n^{2} \rangle = 1 - \frac{n}{N}.
\end{equation}
We next compute the recursion coefficients \(\{R_n,S_n\}\). Although the recursion coefficients can be obtained from the discrete string equations, deriving them directly from Krylov dynamics provides an alternative and conceptually new perspective. For the quadratic potential \(V(\lambda)=\lambda^{2}/2\), the survival amplitude~\eqref{survivalamplitude} evaluates to
\begin{equation}
S(t)
= \frac{1}{h_0}\int e^{-i\lambda t}\, e^{-N\lambda^{2}/2}\, d\lambda
= e^{-t^{2}/(2N)},
\end{equation}
which matches the generating function of the Heisenberg-Weyl algebra~\cite{Caputa:2021sib}. Because the potential is an even function, all odd moments vanish, while the even moments are given by
\begin{equation}
M_{2k}
= \frac{1}{h_0}\int \lambda^{2k}\, d\mu(\lambda)
= \frac{(2k)!}{2^{k} N^{k} k!}.
\end{equation}
Substituting these moments into the recurrence algorithm~\eqref{OCF2} yields
\begin{equation}\label{snrngue}
S_n = 0,
\qquad
R_n = \frac{n}{N},
\end{equation}
in agreement with the discrete string equations~\eqref{VSRREEQ}. In the continuum limit, Eqs.~\eqref{bnensve} and~\eqref{snrngue} give
\begin{equation}
S(x)=a(x)=0,
\qquad
b(x)^2 = R(1-x)=1-x,
\end{equation}
matching the correspondence~\eqref{mainresu1}. Inserting these results into either~\eqref{leadingspectraldensity} or~\eqref{densityofstatesandtheLanczoscoefficients} reproduces the Wigner semicircle,
\begin{equation}
\rho_0(\lambda)
= \frac{1}{\pi} \int_{0}^{1} dx\,
\frac{
\Theta(4x - \lambda^{2})
}{
\sqrt{4x - \lambda^{2}}
}
= \frac{\sqrt{4 - \lambda^{2}}}{2\pi}.
\end{equation}
This recovery of the semicircle law reflects the exact solvability of the GUE, with both the recursion coefficients and the orthogonal polynomials known in closed form. The monic orthogonal polynomials associated with the GUE weight \(e^{-N\lambda^{2}/2}\) follow from Heine’s formula~\eqref{Heineformula} and are given by rescaled Hermite polynomials \cite{Muck:2022xfc},
\begin{equation}\label{PNHERM}
P_n(\lambda)
= (2N)^{-n/2}\,
H_n\!\left(\sqrt{\tfrac{N}{2}}\,\lambda\right),
\qquad
H_n(\lambda)
= (-1)^n e^{\lambda^{2}}
  \frac{d^{\,n}}{d\lambda^{n}} e^{-\lambda^{2}} .
\end{equation}
These polynomials satisfy the recursion relation
\begin{equation}
\lambda\, P_n(\lambda)
= P_{n+1}(\lambda) + \frac{n}{N}\, P_{n-1}(\lambda),
\end{equation}
which is consistent with \(S_n = 0\) and \(R_n = n/N\) as obtained above. Moreover, Eq.~\eqref{psieq} yields a closed-form expression for the Krylov amplitudes \cite{Caputa:2021sib, Muck:2022xfc},
\begin{equation}\label{psintkrywav}
\psi_n(t)
= e^{-t^{2}/(2N)}\, \frac{(-i t)^n}{N^{n/2}\sqrt{n!}},
\qquad
\sum_{n=0}^{\infty} |\psi_n(t)|^{2} = 1 .
\end{equation}
We can substitute Eqs.~\eqref{PNHERM} and~\eqref{psintkrywav} into Eq.~\eqref{krylovdyna} to verify consistency:
\begin{equation}
\begin{aligned}
|\psi(t)\rangle 
&= \sum_{n=0}^\infty \psi_n(t)\, |n\rangle \\
&= \sum_{n=0}^\infty 
e^{-t^2/(2N)}  \frac{(-i t)^n}{N^{n/2} \sqrt{n!}} 
\sqrt{\frac{\sqrt{N}}{2^n n! \sqrt{2\pi}}} \,
H_n\!\left( \sqrt{\tfrac{N}{2}}\, \lambda \right) \\
&= e^{-t^2/(2N)}  \frac{N^{1/4}}{(2\pi)^{1/4}} \,
\exp\left( -i \lambda t + \frac{t^2}{2N} \right) \\
&= \frac{N^{1/4}}{(2\pi)^{1/4}} \, e^{-i \lambda t} \,,
\end{aligned}
\end{equation}
where we used the generating function for Hermite polynomials
\begin{equation}
    \exp(2xy - y^2) = \sum_{n=0}^\infty \frac{H_n(x)\, y^n}{n!} \,.
\end{equation}
Thus, we recover the exact time-evolved quantum state:
\begin{equation}
|\psi(t)\rangle = e^{-i\hat\lambda t} |0\rangle 
= \frac{1}{\sqrt{h_0}}\, e^{-i\lambda t}
= \frac{N^{1/4}}{(2\pi)^{1/4}}\, e^{-i\lambda t} \,.
\end{equation}
Finally, the spread complexity follows directly from the known Krylov amplitudes \eqref{psintkrywav}:
\begin{equation}
C(t) = \sum_{n=0}^\infty n\, |\psi_n(t)|^2 
= \sum_{n=0}^\infty n\, e^{-t^2/N}  \frac{t^{2n}}{N^n n!}
= \frac{t^2}{N} \,.
\end{equation}
The growth of spread complexity at early times is universally quadratic, \(C(t) \sim t^2\), and in the GUE continues to grow quadratically for all times due to the Gaussian property of the measure.

\section{Conclusion and Outlook}\label{sec5}
In this work, we demonstrated that the average Lanczos coefficients $\{b_n, a_n\}$ in the Lanczos approach coincide with the recursion coefficients $\{R_n, S_n\}$ of the orthogonal polynomials approach. In the large-$N$ limit this correspondence becomes
\begin{equation}
    \sqrt{R(x)} = b(1-x)\,, \qquad S(x) = a(1-x)\,,
\end{equation}
and yields identical expressions for the leading density of states in both descriptions. We then defined a Krylov dynamics directly corresponding to the recursion coefficients $\{R_n, S_n\}$, so that the orthogonal polynomials can be interpreted as Krylov polynomials generated by the measure $d\mu(\lambda)$. In this unified framework, quantities such as the survival amplitude, the moments, the recursion (or Lanczos) coefficients, and the spread complexity can be computed using either the Lanczos approach or the orthogonal polynomials approach. We illustrated this framework using the Gaussian Unitary Ensemble, for which the recursion coefficients and orthogonal polynomials are known in closed form. In this example, we verified the agreement of the leading density of states, the Krylov amplitudes, and the resulting complexity growth.

A natural question is whether the auxiliary Hilbert space that appears in the orthogonal polynomials approach has a physical meaning. In double-scaled random matrix models, which have been used to study minimal string theories, the recursion coefficients $\{R(x), S(x)\}$ play a central role \cite{Banks:1989df, Seiberg:2004at}. In parallel, the average Lanczos coefficients $\{b_n, a_n\}$ have been linked to the dynamics of the Einstein–Rosen bridge in JT gravity \cite{Balasubramanian:2024lqk,Nandy:2024zcd}. Given the mapping established in Eq.~\eqref{mainresu1}, it is worth asking whether the auxiliary Hilbert space based on $\{R(x), S(x)\}$ captures emergent geometric or gravitational structure. We expect that exploring this connection and its implications for holography, chaos, and quantum gravity will be a productive direction for future work.

\acknowledgments
It is a pleasure to thank Hong-Yue Jiang for initial collaboration and many useful discussions on this project. I also thank Pawel Caputa, Ben Craps and Yu-Xiao Liu for valuable correspondence. In particular, I would like to thank Shan-Ming Ruan and my supervisor, Juan F.~Pedraza, for carefully reading the manuscript and for many helpful comments and suggestions on the draft. I gratefully acknowledge the financial support provided by the Chinese Scholarship Council (CSC) through a graduate scholarship. I acknowledge the support of the Spanish Agencia Estatal de Investigacion through the grant “IFT Centro de Excelencia Severo Ochoa CEX2020-001007-S”. I am supported through the grants CEX2020-001007-S, PID2021-123017NB-I00 and PID2024-156043NB-I00, funded by MCIN/AEI/10.13039/501100011033, and ERDF, EU.

\appendix
\section{{Non-Gaussian ensembles}}
\label{app:nongaussian}

In this appendix, following \cite{Balasubramanian:2022dnj}, we consider two non-Gaussian ensembles and verify the correspondence between the leading density of states and both the average Lanczos coefficients and the recursion coefficients.

\subsection{Even non-Gaussian ensemble}
We first consider a ensemble whose leading density of states is given by
\begin{equation}\label{leadingden2}
\rho_0(\lambda)
= \frac{\sqrt{4 - \lambda^{2}}}{\pi}
\left(
\frac{7}{10}
- \frac{3}{5} \lambda^{2}
+ \frac{1}{5}\lambda^{4}
\right),
\end{equation}
with support on $\lambda\in[-2,2]$. The corresponding potential is determined via the Coulomb gas method reviewed in \cite{Eynard:2015aea} through the principal value integral
\begin{equation}\label{coulombgas}
\frac{1}{2}\, V'(\lambda)
= \mathrm{p.v.} \int dE \, \frac{\rho_0(E)}{\lambda - E}.
\end{equation}
Evaluating the integral yields the potential
\begin{equation}
V(\lambda)
= \frac{3}{2} \lambda^{2}
- \frac{1}{2}\lambda^{4}
+ \frac{1}{15} \lambda^{6},
\end{equation}
up to an irrelevant additive constant. The saddle-point equations \eqref{extremaofthepotential} associated with this potential reduce to
\begin{equation}
a(x)=0,
\qquad
3 b(x)^{2} - 6 b(x)^{4} + 4 b(x)^{6} = 1 - x .
\end{equation}
Solving the nontrivial branch relevant for the one-cut solution, we obtain
\begin{equation}
a(x)=0,
\qquad
b(x)=\sqrt{\frac{1 + \sqrt[3]{\,1 - 2x\,}}{2}} .
\end{equation}
The recursion coefficients are then determined from the discrete string equations
\eqref{firstdiscretestringequation} and \eqref{seconddiscretestringequation}, leading to
\begin{equation}
S(x)=a(x)=0,
\qquad
\sqrt{R(x)}=b(1-x)
=\sqrt{\frac{1 + \sqrt[3]{\,2x - 1\,}}{2}} ,
\end{equation}
which precisely matches the correspondence proposed in \eqref{mainresu1}. Substituting these expressions into either Eq.~\eqref{leadingspectraldensity} or Eq.~\eqref{densityofstatesandtheLanczoscoefficients},we find
\begin{equation}
\begin{aligned}
\rho_0(\lambda)
&= \frac{1}{\pi} \int_{0}^{1} dx\,
\frac{\Theta\!\left(2 - \lambda^{2} + 2 \sqrt[3]{\,1 - 2x\,}\right)}
{ \sqrt{\,2 - \lambda^{2} + 2 \sqrt[3]{\,1 - 2x\,}}} \\
&= \frac{\sqrt{4 - \lambda^{2}}}{\pi}
\left(
\frac{7}{10}
- \frac{3}{5} \lambda^{2}
+ \frac{1}{5}\lambda^{4}
\right),
\end{aligned}
\end{equation}
thereby exactly reproducing the leading density of states in
Eq.~\eqref{leadingden2}.

\subsection{Non-even non-Gaussian ensemble}
\begin{figure}[t]
  \centering
  \subfigure[]{
    \includegraphics[height=4.6cm,keepaspectratio]{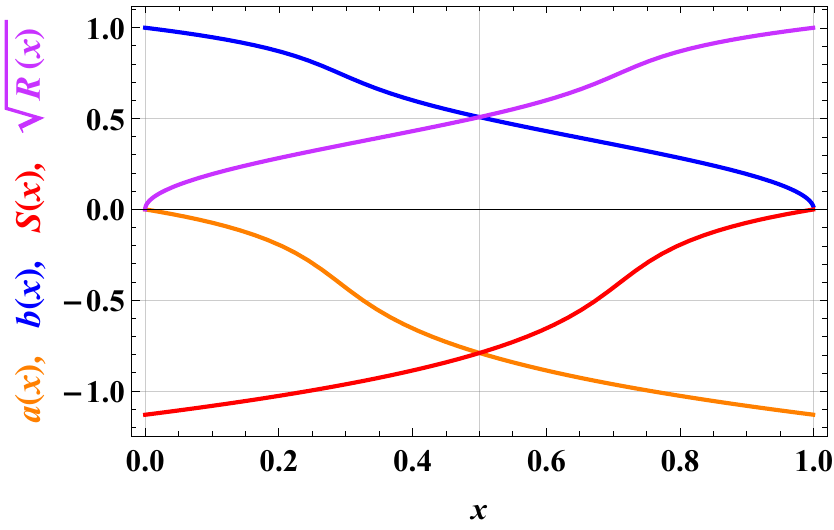}
    \label{RMTN2fig1}
  }
  \subfigure[]{
    \includegraphics[height=4.6cm,keepaspectratio]{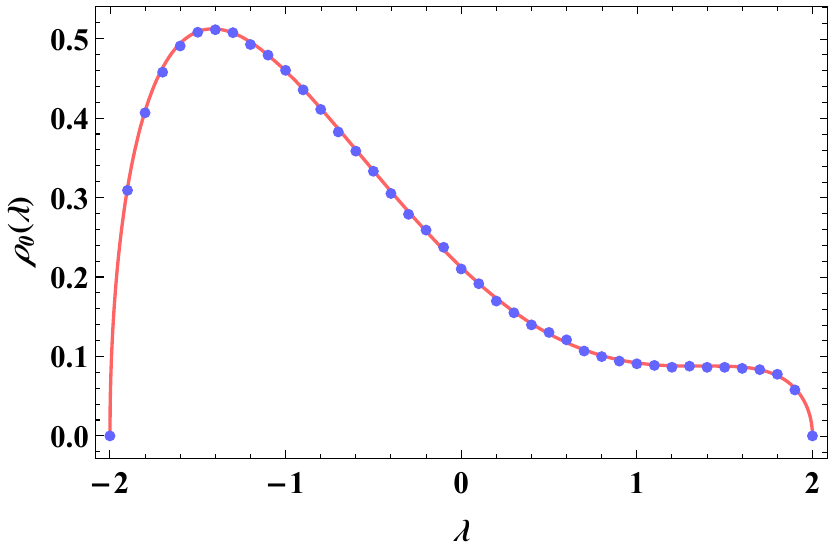}
    \label{RMTN2fig2}
  }
  \caption{
  (a) Average Lanczos coefficients $\{a(x), b(x)\}$, together with the recursion coefficients $\{S(x), \sqrt{R(x)}\}$, as functions of $x$.
  (b) Leading density of states $\rho_0(\lambda)$ as a function of $\lambda$.
  The blue dots denote the numerical results obtained from Eq.~\eqref{leadingspectraldensity} or Eq.~\eqref{densityofstatesandtheLanczoscoefficients},
  while the red curve corresponds to the analytic expression in Eq.~\eqref{leadingden3}.
  }
  \label{RMTN2fig}
\end{figure}
We next consider a ensemble with a non-even leading density of states,
\begin{equation}\label{leadingden3}
\rho_0(\lambda)
= \frac{\sqrt{4 - \lambda^{2}}}{\pi}
\left(
\frac{1}{3}
- \frac{1}{3} \lambda
+ \frac{1}{6}\lambda^{2}
\right).
\end{equation}
According to the relation \eqref{coulombgas}, this density arises from a matrix model with a quartic potential supplemented by cubic and linear terms,
\begin{equation} V(\lambda) = \frac{1}{12} \lambda^{4} - \frac{2}{9}\lambda^{3} + \frac{4}{3} \lambda . \end{equation}
The average Lanczos coefficients satisfy the same saddle-point equations \eqref{extremaofthepotential} as before,
\begin{equation}
\begin{aligned}
0 &= a^3 + 6 a b^2 - 2 a^2 - 4 b^2 + 4 ,\\
1 - x &= b^4 + a^2 b^2 - \frac{4}{3} a b^{2} .
\end{aligned}
\end{equation}
The corresponding recursion coefficients are again governed by the discrete string equations
\eqref{firstdiscretestringequation} and \eqref{seconddiscretestringequation},
\begin{equation}
\begin{aligned}
0 &= S^{3} + 6 S R - 2 S^{2} - 4 R + 4 ,\\
x &= R^{2} + S^{2} R - \frac{4}{3} S R .
\end{aligned}
\end{equation}
These algebraic equations can be solved numerically. The solution is consistent with the correspondence \eqref{mainresu1}, as illustrated in Fig.~\ref{RMTN2fig1}. Substituting the numerical solution into
Eq.~\eqref{leadingspectraldensity} or Eq.~\eqref{densityofstatesandtheLanczoscoefficients} and performing the numerical integration, we recover the leading density of states. The resulting spectral density is in precise agreement with Eq.~\eqref{leadingden3}, as shown in Fig.~\ref{RMTN2fig2}.

\newpage
\bibliography{Refs}

@article{BESSIS1980109,
title = {Quantum field theory techniques in graphical enumeration},
journal = {Advances in Applied Mathematics},
volume = {1},
number = {2},
pages = {109-157},
year = {1980},
issn = {0196-8858},
doi = {https://doi.org/10.1016/0196-8858(80)90008-1},
url = {https://www.sciencedirect.com/science/article/pii/0196885880900081},
author = {D Bessis and C Itzykson and J.B Zuber}
}

@inproceedings{Ginsparg:1991bi,
    author = "Ginsparg, Paul H.",
    title = "{Matrix models of 2-d gravity}",
    eprint = "hep-th/9112013",
    archivePrefix = "arXiv",
    reportNumber = "LA-UR-91-9999, LA-UR-91-4101",
    month = "12",
    year = "1991"
}

@article{Baiguera:2025dkc,
    author = "Baiguera, Stefano and Balasubramanian, Vijay and Caputa, Pawel and Chapman, Shira and Haferkamp, Jonas and Heller, Michal P. and Halpern, Nicole Yunger",
    title = "{Quantum complexity in gravity, quantum field theory, and quantum information science}",
    eprint = "2503.10753",
    archivePrefix = "arXiv",
    primaryClass = "hep-th",
    reportNumber = "YITP-25-39",
    doi = "10.1016/j.physrep.2025.11.001",
    journal = "Phys. Rept.",
    volume = "1159",
    pages = "1--77",
    year = "2026"
}

@article{Huh:2024ytz,
    author = "Huh, Kyoung-Bum and Jeong, Hyun-Sik and Pando Zayas, Leopoldo A. and Pedraza, Juan F.",
    title = "{Krylov complexity in mixed phase space}",
    eprint = "2412.04963",
    archivePrefix = "arXiv",
    primaryClass = "hep-th",
    reportNumber = "LCTP-24-21, IFT-UAM/CSIC-24-170",
    doi = "10.1103/gmy7-dn7l",
    journal = "Phys. Rev. D",
    volume = "111",
    number = "12",
    pages = "L121902",
    year = "2025"
}

@article{Baggioli:2024wbz,
    author = "Baggioli, Matteo and Huh, Kyoung-Bum and Jeong, Hyun-Sik and Kim, Keun-Young and Pedraza, Juan F.",
    title = "{Krylov complexity as an order parameter for quantum chaotic-integrable transitions}",
    eprint = "2407.17054",
    archivePrefix = "arXiv",
    primaryClass = "hep-th",
    reportNumber = "IFT-UAM/CSIC-24-107",
    doi = "10.1103/PhysRevResearch.7.023028",
    journal = "Phys. Rev. Res.",
    volume = "7",
    number = "2",
    pages = "023028",
    year = "2025"
}

@article{Rabinovici:2025otw,
    author = "Rabinovici, Eliezer and S{\'a}nchez-Garrido, Adri{\'a}n and Shir, Ruth and Sonner, Julian",
    title = "{Krylov Complexity}",
    eprint = "2507.06286",
    archivePrefix = "arXiv",
    primaryClass = "hep-th",
    reportNumber = "CERN-TH-2025-128",
    month = "7",
    year = "2025"
}

@article{Nandy:2024zcd,
    author = "Nandy, Pratik",
    title = "{Tridiagonal Hamiltonians modeling the density of states of the double-scaled SYK model}",
    eprint = "2410.07847",
    archivePrefix = "arXiv",
    primaryClass = "hep-th",
    reportNumber = "RIKEN-iTHEMS-Report-24",
    doi = "10.1007/JHEP01(2025)072",
    journal = "JHEP",
    volume = "01",
    pages = "072",
    year = "2025"
}

@article{Balasubramanian:2024lqk,
    author = "Balasubramanian, Vijay and Magan, Javier M. and Nandi, Poulami and Wu, Qingyue",
    title = "{Spread complexity and the saturation of wormhole size}",
    eprint = "2412.02038",
    archivePrefix = "arXiv",
    primaryClass = "hep-th",
    month = "12",
    year = "2024"
}

@article{Banks:1989df,
    author = "Banks, Tom and Douglas, Michael R. and Seiberg, Nathan and Shenker, Stephen H.",
    editor = "Brezin, E. and Wadia, S. R.",
    title = "{Microscopic and Macroscopic Loops in Nonperturbative Two-dimensional Gravity}",
    reportNumber = "RU-89-50",
    doi = "10.1016/0370-2693(90)91736-U",
    journal = "Phys. Lett. B",
    volume = "238",
    pages = "279",
    year = "1990"
}

@article{Seiberg:2004at,
    author = "Seiberg, Nathan and Shih, David",
    title = "{Minimal string theory}",
    eprint = "hep-th/0409306",
    archivePrefix = "arXiv",
    doi = "10.1016/j.crhy.2004.12.007",
    journal = "Comptes Rendus Physique",
    volume = "6",
    pages = "165--174",
    year = "2005"
}

@inproceedings{Ginsparg:1993is,
    author = "Ginsparg, Paul H. and Moore, Gregory W.",
    title = "{Lectures on 2-D gravity and 2-D string theory}",
    booktitle = "{Theoretical Advanced Study Institute (TASI 92): From Black Holes and Strings to Particles}",
    eprint = "hep-th/9304011",
    archivePrefix = "arXiv",
    reportNumber = "YCTP-P23-92, LA-UR-92-3479",
    pages = "277--469",
    month = "10",
    year = "1993"
}

@article{dumitriu2002matrix,
  title={Matrix models for beta ensembles},
  author={Dumitriu, Ioana and Edelman, Alan},
  journal={arXiv preprint math-ph/0206043},
  year={2002}
}

@article{Balasubramanian:2025xkj,
    author = "Balasubramanian, Vijay and Caputa, Pawel and Sim{\'o}n, Joan",
    title = "{Variations on a Theme of Krylov}",
    eprint = "2511.03775",
    archivePrefix = "arXiv",
    primaryClass = "hep-th",
    reportNumber = "YITP-25-171",
    month = "11",
    year = "2025"
}

@article{Caputa:2025ozd,
    author = "Caputa, Pawel and Di Giulio, Giuseppe and Loc, Tran Quang",
    title = "{Symmetry-Resolved Spread Complexity}",
    eprint = "2509.12992",
    archivePrefix = "arXiv",
    primaryClass = "hep-th",
    reportNumber = "YITP-25-146",
    month = "9",
    year = "2025"
}

@article{Dalley:1991jp,
    author = "Dalley, Simon and Johnson, Clifford V. and Morris, Tim R.",
    title = "{Factorization properties of critical matrix models}",
    reportNumber = "SHEP-90-91-19",
    doi = "10.1016/0370-2693(91)90636-5",
    journal = "Phys. Lett. B",
    volume = "262",
    pages = "18--24",
    year = "1991"
}

@article{Dalley:1990zs,
    author = "Dalley, Simon and Johnson, Clifford V. and Morris, Tim R.",
    title = "{Classification of critical hermitian matrix models}",
    reportNumber = "SHEP-90-91-5",
    doi = "10.1142/S0217732391000440",
    journal = "Mod. Phys. Lett. A",
    volume = "6",
    pages = "439--448",
    year = "1991"
}

@article{Aguilar-Gutierrez:2024nau,
    author = "Aguilar-Gutierrez, Sergio E.",
    title = "{Towards complexity in de Sitter space from the doubled-scaled Sachdev-Ye-Kitaev model}",
    eprint = "2403.13186",
    archivePrefix = "arXiv",
    primaryClass = "hep-th",
    doi = "10.1007/JHEP10(2024)107",
    journal = "JHEP",
    volume = "10",
    pages = "107",
    year = "2024"
}

@article{Fu:2025kkh,
    author = "Fu, Yichao and Jeong, Hyun-Sik and Kim, Keun-Young and Pedraza, Juan F.",
    title = "{Toward Krylov-based holography in double-scaled SYK}",
    eprint = "2510.22658",
    archivePrefix = "arXiv",
    primaryClass = "hep-th",
    reportNumber = "IFT-UAM/CSIC-25-105, APCTP Pre2025 - 020",
    month = "10",
    year = "2025"
}

@article{Ambrosini:2025hvo,
    author = "Ambrosini, Marco and Rabinovici, Eliezer and Sonner, Julian",
    title = "{Holography of K-complexity: Switchbacks and Shockwaves}",
    eprint = "2510.17975",
    archivePrefix = "arXiv",
    primaryClass = "hep-th",
    reportNumber = "CERN-TH-2025-206",
    month = "10",
    year = "2025"
}

@article{Xu:2024gfm,
    author = "Xu, Jiuci",
    title = "{On chord dynamics and complexity growth in double-scaled SYK}",
    eprint = "2411.04251",
    archivePrefix = "arXiv",
    primaryClass = "hep-th",
    doi = "10.1007/JHEP06(2025)259",
    journal = "JHEP",
    volume = "06",
    pages = "259",
    year = "2025"
}

@article{Heller:2024ldz,
    author = "Heller, Michal P. and Papalini, Jacopo and Schuhmann, Tim",
    title = "{Krylov Spread Complexity as Holographic Complexity beyond Jackiw-Teitelboim Gravity}",
    eprint = "2412.17785",
    archivePrefix = "arXiv",
    primaryClass = "hep-th",
    doi = "10.1103/spcr-jgm6",
    journal = "Phys. Rev. Lett.",
    volume = "135",
    number = "15",
    pages = "151602",
    year = "2025"
}

@article{He:2025guu,
    author = "He, Peng-Zhang and Liu, Lei-Hua and Zhang, Hai-Qing and Jiang, Qing-Quan",
    title = {{Krylov complexity and Wightman power spectrum with positive chemical potentials in Schr{\"o}dinger field theory}},
    eprint = "2509.14742",
    archivePrefix = "arXiv",
    primaryClass = "hep-th",
    month = "9",
    year = "2025"
}

@article{Caputa:2024sux,
    author = "Caputa, Pawel and Chen, Bowen and McDonald, Ross W. and Sim{\'o}n, Joan and Strittmatter, Benjamin",
    title = "{Spread Complexity Rate as Proper Momentum}",
    eprint = "2410.23334",
    archivePrefix = "arXiv",
    primaryClass = "hep-th",
    reportNumber = "YITP-24-137",
    month = "10",
    year = "2024"
}

@article{Bhattacharya:2024szw,
    author = "Bhattacharya, Aranya and Jana, Aneek",
    title = "{Quantum chaos and complexity from string scattering amplitudes}",
    eprint = "2408.11096",
    archivePrefix = "arXiv",
    primaryClass = "hep-th",
    month = "8",
    year = "2024"
}

@inproceedings{Myers:2024vve,
    author = "Myers, Robert C. and Ruan, Shan-Ming",
    title = "{Complexity Equals (Almost) Anything}",
    eprint = "2403.17475",
    archivePrefix = "arXiv",
    primaryClass = "hep-th",
    reportNumber = "YITP-24-34",
    month = "3",
    year = "2024"
}

@article{Qu:2021ius,
    author = "Qu, Le-Chen and Chen, Jing and Liu, Yu-Xiao",
    title = "{Chaos and complexity for inverted harmonic oscillators}",
    eprint = "2111.07351",
    archivePrefix = "arXiv",
    primaryClass = "hep-th",
    doi = "10.1103/PhysRevD.105.126015",
    journal = "Phys. Rev. D",
    volume = "105",
    number = "12",
    pages = "126015",
    year = "2022"
}

@article{Evnin:2025cfx,
    author = "Evnin, Oleg",
    title = "{Analytic and numerical toolkit for the Anderson model in one dimension}",
    eprint = "2507.06903",
    archivePrefix = "arXiv",
    primaryClass = "cond-mat.dis-nn",
    doi = "10.1103/nmtt-8r72",
    journal = "Phys. Rev. B",
    volume = "112",
    number = "18",
    pages = "L180203",
    year = "2025"
}

@article{Fatemiabhari:2025poq,
    author = "Fatemiabhari, Ali and Nastase, Horatiu and Nunez, Carlos and Roychowdhury, Dibakar",
    title = "{Holographic Krylov Complexity for Conformal Quiver Gauge Theories}",
    eprint = "2512.14812",
    archivePrefix = "arXiv",
    primaryClass = "hep-th",
    month = "12",
    year = "2025"
}

@article{Fatemiabhari:2025cyy,
    author = "Fatemiabhari, Ali and Nastase, Horatiu and Roychowdhury, Dibakar",
    title = "{Holographic Krylov complexity in ${\cal N}=4$ SYM}",
    eprint = "2511.19286",
    archivePrefix = "arXiv",
    primaryClass = "hep-th",
    month = "11",
    year = "2025"
}

@article{Fatemiabhari:2025usn,
    author = "Fatemiabhari, Ali and Nastase, Horatiu and Nunez, Carlos and Roychowdhury, Dibakar",
    title = "{Holographic Krylov complexity in confining gauge theories}",
    eprint = "2511.22717",
    archivePrefix = "arXiv",
    primaryClass = "hep-th",
    month = "11",
    year = "2025"
}

@article{Qu:2022zwq,
    author = "Qu, Le-Chen and Jiang, Hong-Yue and Liu, Yu-Xiao",
    title = "{Chaos and multifold complexity for an inverted harmonic oscillator}",
    eprint = "2211.04317",
    archivePrefix = "arXiv",
    primaryClass = "quant-ph",
    doi = "10.1007/JHEP12(2022)065",
    journal = "JHEP",
    volume = "12",
    pages = "065",
    year = "2022"
}

@article{Caceres:2025myu,
    author = "C{\'a}ceres, Elena and Carrasco, Rafael and Patil, Vaishnavi and Pedraza, Juan F. and Svesko, Andrew",
    title = "{The landscape of complexity measures in 2D gravity}",
    eprint = "2503.20943",
    archivePrefix = "arXiv",
    primaryClass = "hep-th",
    reportNumber = "IFT-UAM/CSIC-25-30, IFT-UAM/CSIC-25-30",
    doi = "10.1007/JHEP10(2025)218",
    journal = "JHEP",
    volume = "10",
    pages = "218",
    year = "2025"
}

@article{Caceres:2025ypk,
    author = "C{\'a}ceres, Elena and Carrasco, Rafael and Pedraza, Juan F.",
    title = "{Lorentzian threads and nonlocal computation in holography}",
    eprint = "2512.07963",
    archivePrefix = "arXiv",
    primaryClass = "hep-th",
    reportNumber = "WI-42-2025, IFT-UAM/CSIC-25-156",
    month = "12",
    year = "2025"
}

@article{Miyaji:2025jxy,
    author = "Miyaji, Masamichi and Ruan, Shan-Ming and Shibuya, Shono and Yano, Kazuyoshi",
    title = "{Universal Time Evolution of Holographic and Quantum Complexity}",
    eprint = "2507.23667",
    archivePrefix = "arXiv",
    primaryClass = "hep-th",
    reportNumber = "RIKEN-iTHEMS-Report-25",
    month = "7",
    year = "2025"
}

@article{Jiang:2025qai,
    author = "Jiang, Hong-Yue and Liu, Yu-Xiao",
    title = "{Complexity equals anything for multi-horizon black holes}",
    eprint = "2506.10398",
    archivePrefix = "arXiv",
    primaryClass = "hep-th",
    doi = "10.1007/JHEP12(2025)072",
    journal = "JHEP",
    volume = "12",
    pages = "072",
    year = "2025"
}

@article{Arean:2024pzo,
    author = "Are{\'a}n, Daniel and Jeong, Hyun-Sik and Pedraza, Juan F. and Qu, Le-Chen",
    title = "{Kasner interiors from analytic hairy black holes}",
    eprint = "2407.18430",
    archivePrefix = "arXiv",
    primaryClass = "hep-th",
    reportNumber = "IFT-UAM/CSIC-24-109",
    doi = "10.1007/JHEP11(2024)138",
    journal = "JHEP",
    volume = "11",
    pages = "138",
    year = "2024"
}

@article{Caceres:2023ziv,
    author = "Caceres, Elena and Carrasco, Rafael and Patil, Vaishnavi",
    title = "{Lorentzian threads and generalized complexity}",
    eprint = "2312.10606",
    archivePrefix = "arXiv",
    primaryClass = "hep-th",
    doi = "10.1007/JHEP04(2024)010",
    journal = "JHEP",
    volume = "04",
    pages = "010",
    year = "2024"
}

@article{Jiang:2023jti,
    author = "Jiang, Hong-Yue and Wang, Meng-Ting and Liua, Yu-Xiao",
    title = "{Holographic complexity and phase transition for AdS black holes}",
    eprint = "2307.09223",
    archivePrefix = "arXiv",
    primaryClass = "hep-th",
    doi = "10.1103/PhysRevD.110.046013",
    journal = "Phys. Rev. D",
    volume = "110",
    number = "4",
    pages = "046013",
    year = "2024"
}

@article{Jorstad:2023kmq,
    author = "J{\o}rstad, Eivind and Myers, Robert C. and Ruan, Shan-Ming",
    title = "{Complexity=anything: singularity probes}",
    eprint = "2304.05453",
    archivePrefix = "arXiv",
    primaryClass = "hep-th",
    reportNumber = "YITP-23-41",
    doi = "10.1007/JHEP07(2023)223",
    journal = "JHEP",
    volume = "07",
    pages = "223",
    year = "2023"
}

@article{Belin:2022xmt,
    author = "Belin, Alexandre and Myers, Robert C. and Ruan, Shan-Ming and S{\'a}rosi, G{\'a}bor and Speranza, Antony J.",
    title = "{Complexity equals anything II}",
    eprint = "2210.09647",
    archivePrefix = "arXiv",
    primaryClass = "hep-th",
    reportNumber = "CERN-TH-2022-159; YITP-22-101",
    doi = "10.1007/JHEP01(2023)154",
    journal = "JHEP",
    volume = "01",
    pages = "154",
    year = "2023"
}

@article{Ruan:2020vze,
    author = "Ruan, Shan-Ming",
    title = "{Purification Complexity without Purifications}",
    eprint = "2006.01088",
    archivePrefix = "arXiv",
    primaryClass = "hep-th",
    doi = "10.1007/JHEP01(2021)092",
    journal = "JHEP",
    volume = "01",
    pages = "092",
    year = "2021"
}

@article{Bernamonti:2020bcf,
    author = "Bernamonti, Alice and Galli, Federico and Hernandez, Juan and Myers, Robert C. and Ruan, Shan-Ming and Sim{\'o}n, Joan",
    title = "{Aspects of The First Law of Complexity}",
    eprint = "2002.05779",
    archivePrefix = "arXiv",
    primaryClass = "hep-th",
    doi = "10.1088/1751-8121/ab8e66",
    journal = "J. Phys. A",
    volume = "53",
    pages = "29",
    year = "2020"
}

@article{Bernamonti:2019zyy,
    author = "Bernamonti, Alice and Galli, Federico and Hernandez, Juan and Myers, Robert C. and Ruan, Shan-Ming and Sim{\'o}n, Joan",
    title = "{First Law of Holographic Complexity}",
    eprint = "1903.04511",
    archivePrefix = "arXiv",
    primaryClass = "hep-th",
    doi = "10.1103/PhysRevLett.123.081601",
    journal = "Phys. Rev. Lett.",
    volume = "123",
    number = "8",
    pages = "081601",
    year = "2019"
}

@article{Demulder:2025uda,
    author = "Demulder, Saskia and Knysh, Maria and Rolph, Andrew",
    title = "{Krylov exponents and power spectra for maximal quantum chaos: an EFT approach}",
    eprint = "2508.05444",
    archivePrefix = "arXiv",
    primaryClass = "hep-th",
    month = "8",
    year = "2025"
}

@article{Miyaji:2025ucp,
    author = "Miyaji, Masamichi and Mori, Soichiro and Okuyama, Kazumi",
    title = "{Finite N bulk Hilbert space in ETH matrix model for double-scaled SYK. Null states, state-dependence and Krylov state complexity}",
    eprint = "2505.13194",
    archivePrefix = "arXiv",
    primaryClass = "hep-th",
    reportNumber = "YITP 25-71",
    doi = "10.1007/JHEP08(2025)084",
    journal = "JHEP",
    volume = "08",
    pages = "084",
    year = "2025"
}

@article{Takahashi:2025iol,
    author = "Takahashi, Kazutaka and Nandy, Pratik and del Campo, Adolfo",
    title = "{Krylov Complexity Under Hamiltonian Deformations and Toda Flows}",
    eprint = "2510.19436",
    archivePrefix = "arXiv",
    primaryClass = "quant-ph",
    reportNumber = "RIKEN-iTHEMS-Report-25",
    month = "10",
    year = "2025"
}

@article{Bhattacharya:2024uxx,
    author = "Bhattacharya, Aranya and Nath, Pingal Pratyush and Sahu, Himanshu",
    title = "{Speed limits to the growth of Krylov complexity in open quantum systems}",
    eprint = "2403.03584",
    archivePrefix = "arXiv",
    primaryClass = "quant-ph",
    doi = "10.1103/PhysRevD.109.L121902",
    journal = "Phys. Rev. D",
    volume = "109",
    number = "12",
    pages = "L121902",
    year = "2024"
}

@article{Bhattacharya:2024hto,
    author = "Bhattacharya, Aranya and Das, Rathindra Nath and Dey, Bidyut and Erdmenger, Johanna",
    title = "{Spread complexity and localization in PT-symmetric systems}",
    eprint = "2406.03524",
    archivePrefix = "arXiv",
    primaryClass = "hep-th",
    doi = "10.1103/PhysRevB.110.064320",
    journal = "Phys. Rev. B",
    volume = "110",
    number = "6",
    pages = "064320",
    year = "2024"
}

@article{Li:2024ljz,
    author = "Li, Tao and Liu, Lei-Hua",
    title = "{Krylov complexity of thermal state in early universe}",
    eprint = "2408.03293",
    archivePrefix = "arXiv",
    primaryClass = "hep-th",
    month = "8",
    year = "2024"
}

@article{Miyaji:2025yvm,
    author = "Miyaji, Masamichi and Ruan, Shan-Ming and Shibuya, Shono and Yano, Kazuyoshi",
    title = "{Non-perturbative overlaps in JT gravity: from spectral form factor to generating functions of complexity}",
    eprint = "2502.12266",
    archivePrefix = "arXiv",
    primaryClass = "hep-th",
    reportNumber = "YITP-25-28",
    doi = "10.1007/JHEP06(2025)251",
    journal = "JHEP",
    volume = "06",
    pages = "251",
    year = "2025"
}

@article{Caputa:2025dep,
    author = "Caputa, Pawel and Di Giulio, Giuseppe",
    title = "{Local quenches from a Krylov perspective}",
    eprint = "2502.19485",
    archivePrefix = "arXiv",
    primaryClass = "hep-th",
    reportNumber = "YITP-25-26",
    doi = "10.1007/JHEP07(2025)164",
    journal = "JHEP",
    volume = "07",
    pages = "164",
    year = "2025"
}

@article{Caputa:2025mii,
    author = "Caputa, Pawel and Di Giulio, Giuseppe and Loc, Tran Quang",
    title = "{Growth of block-diagonal operators and symmetry-resolved Krylov complexity}",
    eprint = "2507.02033",
    archivePrefix = "arXiv",
    primaryClass = "hep-th",
    reportNumber = "YITP-25-101",
    doi = "10.1103/9v9v-54zv",
    journal = "Phys. Rev. Res.",
    volume = "7",
    number = "4",
    pages = "043055",
    year = "2025"
}

@article{Balasubramanian:2024ghv,
    author = "Balasubramanian, Vijay and Das, Rathindra Nath and Erdmenger, Johanna and Xian, Zhuo-Yu",
    title = "{Chaos and integrability in triangular billiards}",
    eprint = "2407.11114",
    archivePrefix = "arXiv",
    primaryClass = "hep-th",
    doi = "10.1088/1742-5468/adba41",
    journal = "J. Stat. Mech.",
    volume = "2025",
    number = "3",
    pages = "033202",
    year = "2025"
}

@article{Nandy:2024mml,
    author = "Nandy, Pratik and Pathak, Tanay and Xian, Zhuo-Yu and Erdmenger, Johanna",
    title = "{Krylov space approach to singular value decomposition in non-Hermitian systems}",
    eprint = "2411.09309",
    archivePrefix = "arXiv",
    primaryClass = "quant-ph",
    reportNumber = "YITP-24-132, RIKEN-iTHEMS-Report-24",
    doi = "10.1103/PhysRevB.111.064203",
    journal = "Phys. Rev. B",
    volume = "111",
    number = "6",
    pages = "064203",
    year = "2025"
}

@article{He:2024xjp,
    author = "He, Peng-Zhang and Zhang, Hai-Qing",
    title = "{Probing Krylov complexity in scalar field theory with general temperatures}",
    eprint = "2407.02756",
    archivePrefix = "arXiv",
    primaryClass = "hep-th",
    doi = "10.1007/JHEP11(2024)014",
    journal = "JHEP",
    volume = "11",
    pages = "014",
    year = "2024"
}

@article{He:2024hkw,
    author = "He, Peng-Zhang and Zhang, Hai-Qing",
    title = {{Krylov complexity in the Schr{\"o}dinger field theory}},
    eprint = "2411.16302",
    archivePrefix = "arXiv",
    primaryClass = "hep-th",
    doi = "10.1007/JHEP03(2025)142",
    journal = "JHEP",
    volume = "03",
    pages = "142",
    year = "2025"
}

@article{Zhai:2024tkz,
    author = "Zhai, Ke-Hong and Liu, Lei-Hua and Zhang, Hai-Qing",
    title = "{The generalized CV conjecture of Krylov complexity}",
    eprint = "2412.08925",
    archivePrefix = "arXiv",
    primaryClass = "hep-th",
    month = "12",
    year = "2024"
}

@article{Zhai:2025abc,
    author = "Zhai, Ke-Hong and Liu, Lei-Hua and Zhang, Hai-Qing",
    title = "{Inflationary power spectrum from the Lanczos algorithm}",
    eprint = "2505.20595",
    archivePrefix = "arXiv",
    primaryClass = "quant-ph",
    doi = "10.1140/epjc/s10052-025-14791-w",
    journal = "Eur. Phys. J. C",
    volume = "85",
    number = "10",
    pages = "1096",
    year = "2025"
}

@article{Berkooz:2018qkz,
    author = "Berkooz, Micha and Narayan, Prithvi and Simon, Joan",
    title = "{Chord diagrams, exact correlators in spin glasses and black hole bulk reconstruction}",
    eprint = "1806.04380",
    archivePrefix = "arXiv",
    primaryClass = "hep-th",
    doi = "10.1007/JHEP08(2018)192",
    journal = "JHEP",
    volume = "08",
    pages = "192",
    year = "2018"
}

@article{Dalley:1990hb,
    author = "Dalley, Simon",
    title = "{Critical conditions for matrix models of string theory}",
    reportNumber = "SHEP-90-91-6",
    month = "11",
    year = "1990"
}

@article{livan2018introduction,
  title={Introduction to random matrices theory and practice},
  author={Livan, Giacomo and Novaes, Marcel and Vivo, Pierpaolo},
  journal={Monograph Award},
  volume={63},
  number={54},
  pages={914},
  year={2018}
}

@incollection{bleher2011lectures,
  title={Lectures on random matrix models: the Riemann--Hilbert approach},
  author={Bleher, Pavel M},
  booktitle={Random Matrices, Random Processes and Integrable Systems},
  pages={251--349},
  year={2011},
  publisher={Springer}
}

@article{Eynard:2015aea,
    author = "Eynard, Bertrand and Kimura, Taro and Ribault, Sylvain",
    title = "{Random matrices}",
    eprint = "1510.04430",
    archivePrefix = "arXiv",
    primaryClass = "math-ph",
    month = "10",
    year = "2015"
}

@article{Adhikari:2025vdl,
    author = "Adhikari, Kiran",
    title = "{Krylov Polynomials and Quantum Query Complexity}",
    eprint = "2510.11786",
    archivePrefix = "arXiv",
    primaryClass = "quant-ph",
    month = "10",
    year = "2025"
}

@article{Alishahiha:2024vbf,
    author = "Alishahiha, Mohsen and Banerjee, Souvik and Vasli, Mohammad Javad",
    title = "{Krylov complexity as a probe for chaos}",
    eprint = "2408.10194",
    archivePrefix = "arXiv",
    primaryClass = "hep-th",
    doi = "10.1140/epjc/s10052-025-14451-z",
    journal = "Eur. Phys. J. C",
    volume = "85",
    number = "7",
    pages = "749",
    year = "2025"
}

@article{Muck:2024fpb,
    author = {M{\"u}ck, Wolfgang},
    title = "{Black holes and Marchenko-Pastur distribution}",
    eprint = "2403.05241",
    archivePrefix = "arXiv",
    primaryClass = "hep-th",
    doi = "10.1103/PhysRevD.109.126001",
    journal = "Phys. Rev. D",
    volume = "109",
    number = "12",
    pages = "126001",
    year = "2024"
}

@article{Craps:2025kub,
    author = "Craps, Ben and Pascuzzi, Gabriele and Pedraza, Juan F. and Qu, Le-Chen and Ruan, Shan-Ming",
    title = "{Explicit Connections Between Krylov and Nielsen Complexity}",
    eprint = "2511.15799",
    archivePrefix = "arXiv",
    primaryClass = "hep-th",
    reportNumber = "IFT-UAM/CSIC-25-123",
    month = "11",
    year = "2025"
}

@article{bleher2008lecturesrandommatrixmodels,
    author = "Pavel M. Bleher",
    title = "{Lectures on random matrix models. The Riemann-Hilbert approach}",
    eprint = "0801.1858",
    archivePrefix = "arXiv",
    primaryClass = "math-ph",
    year = "2008"
}

@article{Craps:2024suj,
    author = "Craps, Ben and Evnin, Oleg and Pascuzzi, Gabriele",
    title = "{Multiseed Krylov complexity}",
    eprint = "2409.15666",
    archivePrefix = "arXiv",
    primaryClass = "quant-ph",
    month = "9",
    year = "2024"
}

@article{Nandy:2024htc,
    author = "Nandy, Pratik and Matsoukas-Roubeas, Apollonas S. and Mart\'\i{}nez-Azcona, Pablo and Dymarsky, Anatoly and del Campo, Adolfo",
    title = "{Quantum Dynamics in Krylov Space: Methods and Applications}",
    eprint = "2405.09628",
    archivePrefix = "arXiv",
    primaryClass = "quant-ph",
    reportNumber = "RIKEN-iTHEMS-Report-24",
    month = "5",
    year = "2024"
}

@article{Dowling:2006tnk,
    author = "Dowling, Mark R. and Nielsen, Michael A.",
    title = "{The geometry of quantum computation}",
    eprint = "quant-ph/0701004",
    archivePrefix = "arXiv",
    doi = "10.26421/QIC8.10-1",
    journal = "Quant. Inf. Comput.",
    volume = "8",
    number = "10",
    pages = "0861--0899",
    year = "2008"
}

@article{Craps:2023ivc,
    author = "Craps, Ben and Evnin, Oleg and Pascuzzi, Gabriele",
    title = "{A Relation between Krylov and Nielsen Complexity}",
    eprint = "2311.18401",
    archivePrefix = "arXiv",
    primaryClass = "quant-ph",
    doi = "10.1103/PhysRevLett.132.160402",
    journal = "Phys. Rev. Lett.",
    volume = "132",
    number = "16",
    pages = "160402",
    year = "2024"
}

@article{Haque:2024ldr,
    author = "Haque, S. Shajidul and Jafari, Ghadir and Underwood, Bret",
    title = "{Universal early-time growth in quantum circuit complexity}",
    eprint = "2406.12990",
    archivePrefix = "arXiv",
    primaryClass = "hep-th",
    doi = "10.1007/JHEP10(2024)101",
    journal = "JHEP",
    volume = "10",
    pages = "101",
    year = "2024"
}

@article{Nielsen:2005mkt,
    author = "Nielsen, Michael A.",
    title = "{A geometric approach to quantum circuit lower bounds}",
    eprint = "quant-ph/0502070",
    archivePrefix = "arXiv",
    doi = "10.26421/QIC6.3-2",
    journal = "Quant. Inf. Comput.",
    volume = "6",
    number = "3",
    pages = "213--262",
    year = "2006"
}

@article{Nielsen:2006cea,
    author = "Nielsen, Michael A. and Dowling, Mark R. and Gu, Mile and Doherty, Andrew C.",
    title = "{Quantum Computation as Geometry}",
    eprint = "quant-ph/0603161",
    archivePrefix = "arXiv",
    doi = "10.1126/science.1121541",
    journal = "Science",
    volume = "311",
    number = "5764",
    pages = "1133--1135",
    year = "2006"
}

@article{Caputa:2024vrn,
    author = "Caputa, Pawel and Jeong, Hyun-Sik and Liu, Sinong and Pedraza, Juan F. and Qu, Le-Chen",
    title = "{Krylov complexity of density matrix operators}",
    eprint = "2402.09522",
    archivePrefix = "arXiv",
    primaryClass = "hep-th",
    reportNumber = "YITP-24-21, IFT-UAM/CSIC-24-25",
    month = "2",
    year = "2024"
}

@article{Loc:2024oen,
	archiveprefix = {arXiv},
	author = {Loc, Tran Quang},
	eprint = {2402.07980},
	month = {2},
	primaryclass = {hep-th},
	title = {{Lanczos spectrum for random operator growth}},
	year = {2024}}

@article{Carrasco:2023fcj,
	archiveprefix = {arXiv},
	author = {Carrasco, Rafael and Pedraza, Juan F. and Svesko, Andrew and Weller-Davies, Zachary},
	doi = {10.1007/JHEP09(2023)167},
	eprint = {2306.08503},
	journal = {JHEP},
	pages = {167},
	primaryclass = {hep-th},
	reportnumber = {IFT-UAM/CSIC-23-71},
	title = {{Gravitation from optimized computation: Einstein and beyond}},
	volume = {09},
	year = {2023}}

@article{Belin:2021bga,
	archiveprefix = {arXiv},
	author = {Belin, Alexandre and Myers, Robert C. and Ruan, Shan-Ming and S\'arosi, G\'abor and Speranza, Antony J.},
	doi = {10.1103/PhysRevLett.128.081602},
	eprint = {2111.02429},
	journal = {Phys. Rev. Lett.},
	number = {8},
	pages = {081602},
	primaryclass = {hep-th},
	reportnumber = {CERN-TH-2021-181, YITP-22-02},
	title = {{Does Complexity Equal Anything?}},
	volume = {128},
	year = {2022}}

@article{Caceres:2019pgf,
	archiveprefix = {arXiv},
	author = {Caceres, Elena and Chapman, Shira and Couch, Josiah D. and Hernandez, Juan P. and Myers, Robert C. and Ruan, Shan-Ming},
	doi = {10.1007/JHEP03(2020)012},
	eprint = {1909.10557},
	journal = {JHEP},
	pages = {012},
	primaryclass = {hep-th},
	title = {{Complexity of Mixed States in QFT and Holography}},
	volume = {03},
	year = {2020}}

@article{Caputa:2023vyr,
	archiveprefix = {arXiv},
	author = {Caputa, Pawel and Magan, Javier M. and Patramanis, Dimitrios and Tonni, Erik},
	eprint = {2306.14732},
	month = {6},
	primaryclass = {hep-th},
	title = {{Krylov complexity of modular Hamiltonian evolution}},
	year = {2023}}

@article{Balasubramanian:2022dnj,
	archiveprefix = {arXiv},
	author = {Balasubramanian, Vijay and Magan, Javier M. and Wu, Qingyue},
	doi = {10.1103/PhysRevD.107.126001},
	eprint = {2208.08452},
	journal = {Phys. Rev. D},
	number = {12},
	pages = {126001},
	primaryclass = {hep-th},
	title = {{Tridiagonalizing random matrices}},
	volume = {107},
	year = {2023}}

@article{Kar:2021nbm,
	archiveprefix = {arXiv},
	author = {Kar, Arjun and Lamprou, Lampros and Rozali, Moshe and Sully, James},
	doi = {10.1007/JHEP01(2022)016},
	eprint = {2106.02046},
	journal = {JHEP},
	pages = {016},
	primaryclass = {hep-th},
	title = {{Random matrix theory for complexity growth and black hole interiors}},
	volume = {01},
	year = {2022}}

@article{Balasubramanian:2023kwd,
	archiveprefix = {arXiv},
	author = {Balasubramanian, Vijay and Magan, Javier M. and Wu, Qingyue},
	eprint = {2312.03848},
	month = {12},
	primaryclass = {hep-th},
	title = {{Quantum chaos, integrability, and late times in the Krylov basis}},
	year = {2023}}

@article{Ali:2019zcj,
	archiveprefix = {arXiv},
	author = {Ali, Tibra and Bhattacharyya, Arpan and Haque, S. Shajidul and Kim, Eugene H. and Moynihan, Nathan and Murugan, Jeff},
	doi = {10.1103/PhysRevD.101.026021},
	eprint = {1905.13534},
	journal = {Phys. Rev. D},
	number = {2},
	pages = {026021},
	primaryclass = {hep-th},
	reportnumber = {YITP-19-45},
	title = {{Chaos and Complexity in Quantum Mechanics}},
	volume = {101},
	year = {2020}}

@article{Susskind:2014rva,
	archiveprefix = {arXiv},
	author = {Susskind, Leonard},
	doi = {10.1002/prop.201500092},
	eprint = {1403.5695},
	journal = {Fortsch. Phys.},
	note = {[Addendum: Fortsch.Phys. 64, 44--48 (2016)]},
	pages = {24--43},
	primaryclass = {hep-th},
	title = {{Computational Complexity and Black Hole Horizons}},
	volume = {64},
	year = {2016}}

@article{Erdmenger:2023wjg,
	archiveprefix = {arXiv},
	author = {Erdmenger, Johanna and Jian, Shao-Kai and Xian, Zhuo-Yu},
	doi = {10.1007/JHEP08(2023)176},
	eprint = {2303.12151},
	journal = {JHEP},
	pages = {176},
	primaryclass = {hep-th},
	title = {{Universal chaotic dynamics from Krylov space}},
	volume = {08},
	year = {2023}}

@article{Avdoshkin:2019trj,
	archiveprefix = {arXiv},
	author = {Avdoshkin, Alexander and Dymarsky, Anatoly},
	doi = {10.1103/PhysRevResearch.2.043234},
	eprint = {1911.09672},
	journal = {Phys. Rev. Res.},
	number = {4},
	pages = {043234},
	primaryclass = {cond-mat.stat-mech},
	title = {{Euclidean operator growth and quantum chaos}},
	volume = {2},
	year = {2020}}

@article{Balasubramanian:2022tpr,
	archiveprefix = {arXiv},
	author = {Balasubramanian, Vijay and Caputa, Pawel and Magan, Javier M. and Wu, Qingyue},
	doi = {10.1103/PhysRevD.106.046007},
	eprint = {2202.06957},
	journal = {Phys. Rev. D},
	number = {4},
	pages = {046007},
	primaryclass = {hep-th},
	title = {{Quantum chaos and the complexity of spread of states}},
	volume = {106},
	year = {2022}}

@article{Dymarsky:2021bjq,
	archiveprefix = {arXiv},
	author = {Dymarsky, Anatoly and Smolkin, Michael},
	doi = {10.1103/PhysRevD.104.L081702},
	eprint = {2104.09514},
	journal = {Phys. Rev. D},
	number = {8},
	pages = {L081702},
	primaryclass = {hep-th},
	title = {{Krylov complexity in conformal field theory}},
	volume = {104},
	year = {2021}}

@article{Hornedal:2022pkc,
	archiveprefix = {arXiv},
	author = {H\"ornedal, Niklas and Carabba, Nicoletta and Matsoukas-Roubeas, Apollonas S. and del Campo, Adolfo},
	eprint = {2202.05006},
	month = {2},
	primaryclass = {quant-ph},
	title = {{Ultimate Physical Limits to the Growth of Operator Complexity}},
	year = {2022}}

@article{Caputa:2021sib,
	archiveprefix = {arXiv},
	author = {Caputa, Pawel and Magan, Javier M. and Patramanis, Dimitrios},
	doi = {10.1103/PhysRevResearch.4.013041},
	eprint = {2109.03824},
	journal = {Phys. Rev. Res.},
	number = {1},
	pages = {013041},
	primaryclass = {hep-th},
	title = {{Geometry of Krylov complexity}},
	volume = {4},
	year = {2022}}

@article{Rabinovici:2020ryf,
	archiveprefix = {arXiv},
	author = {Rabinovici, E. and S\'anchez-Garrido, A. and Shir, R. and Sonner, J.},
	doi = {10.1007/JHEP06(2021)062},
	eprint = {2009.01862},
	journal = {JHEP},
	pages = {062},
	primaryclass = {hep-th},
	title = {{Operator complexity: a journey to the edge of Krylov space}},
	volume = {06},
	year = {2021}}

@article{Jian:2020qpp,
	archiveprefix = {arXiv},
	author = {Jian, Shao-Kai and Swingle, Brian and Xian, Zhuo-Yu},
	doi = {10.1007/JHEP03(2021)014},
	eprint = {2008.12274},
	journal = {JHEP},
	pages = {014},
	primaryclass = {hep-th},
	title = {{Complexity growth of operators in the SYK model and in JT gravity}},
	volume = {03},
	year = {2021}}

@article{Muck:2022xfc,
	archiveprefix = {arXiv},
	author = {M\"uck, Wolfgang and Yang, Yi},
	doi = {10.1016/j.nuclphysb.2022.115948},
	eprint = {2205.12815},
	journal = {Nucl. Phys. B},
	pages = {115948},
	primaryclass = {hep-th},
	title = {{Krylov complexity and orthogonal polynomials}},
	volume = {984},
	year = {2022}}

@article{Barbon:2019wsy,
	archiveprefix = {arXiv},
	author = {Barb\'on, J. L. F. and Rabinovici, E. and Shir, R. and Sinha, R.},
	doi = {10.1007/JHEP10(2019)264},
	eprint = {1907.05393},
	journal = {JHEP},
	pages = {264},
	primaryclass = {hep-th},
	reportnumber = {IFT-UAM/CSIC-19-98},
	title = {{On The Evolution Of Operator Complexity Beyond Scrambling}},
	volume = {10},
	year = {2019}}

@article{Parker:2018yvk,
	archiveprefix = {arXiv},
	author = {Parker, Daniel E. and Cao, Xiangyu and Avdoshkin, Alexander and Scaffidi, Thomas and Altman, Ehud},
	doi = {10.1103/PhysRevX.9.041017},
	eprint = {1812.08657},
	journal = {Phys. Rev. X},
	number = {4},
	pages = {041017},
	primaryclass = {cond-mat.stat-mech},
	title = {{A Universal Operator Growth Hypothesis}},
	volume = {9},
	year = {2019}}

@article{Rabinovici:2023yex,
	archiveprefix = {arXiv},
	author = {Rabinovici, E. and S\'anchez-Garrido, A. and Shir, R. and Sonner, J.},
	doi = {10.1007/JHEP08(2023)213},
	eprint = {2305.04355},
	journal = {JHEP},
	pages = {213},
	primaryclass = {hep-th},
	title = {{A bulk manifestation of Krylov complexity}},
	volume = {08},
	year = {2023}}

@article{Lin:2022rbf,
	archiveprefix = {arXiv},
	author = {Lin, Henry W.},
	doi = {10.1007/JHEP11(2022)060},
	eprint = {2208.07032},
	journal = {JHEP},
	pages = {060},
	primaryclass = {hep-th},
	title = {{The bulk Hilbert space of double scaled SYK}},
	volume = {11},
	year = {2022}}

@article{Chapman:2018hou,
	archiveprefix = {arXiv},
	author = {Chapman, Shira and Eisert, Jens and Hackl, Lucas and Heller, Michal P. and Jefferson, Ro and Marrochio, Hugo and Myers, Robert C.},
	eprint = {1810.05151},
	primaryclass = {hep-th},
	slaccitation = {%%CITATION = ARXIV:1810.05151;%%},
	title = {{Complexity and entanglement for thermofield double states}},
	year = {2018}}

@article{Bhattacharyya:2018bbv,
	archiveprefix = {arXiv},
	author = {Bhattacharyya, Arpan and Shekar, Arvind and Sinha, Aninda},
	doi = {10.1007/JHEP10(2018)140},
	eprint = {1808.03105},
	journal = {JHEP},
	pages = {140},
	primaryclass = {hep-th},
	reportnumber = {YITP-18-89},
	slaccitation = {%%CITATION = ARXIV:1808.03105;%%},
	title = {{Circuit complexity in interacting QFTs and RG flows}},
	volume = {10},
	year = {2018}}

@article{Guo:2018kzl,
	archiveprefix = {arXiv},
	author = {Guo, Minyong and Hernandez, Juan and Myers, Robert C. and Ruan, Shan-Ming},
	doi = {10.1007/JHEP10(2018)011},
	eprint = {1807.07677},
	journal = {JHEP},
	pages = {011},
	primaryclass = {hep-th},
	slaccitation = {%%CITATION = ARXIV:1807.07677;%%},
	title = {{Circuit Complexity for Coherent States}},
	volume = {10},
	year = {2018}}

@article{Hackl:2018ptj,
	archiveprefix = {arXiv},
	author = {Hackl, Lucas and Myers, Robert C.},
	doi = {10.1007/JHEP07(2018)139},
	eprint = {1803.10638},
	journal = {JHEP},
	pages = {139},
	primaryclass = {hep-th},
	reportnumber = {IGC-18/3-1, IGC-18-3-1},
	slaccitation = {%%CITATION = ARXIV:1803.10638;%%},
	title = {{Circuit complexity for free fermions}},
	volume = {07},
	year = {2018}}

@article{Khan:2018rzm,
	archiveprefix = {arXiv},
	author = {Khan, Rifath and Krishnan, Chethan and Sharma, Sanchita},
	doi = {10.1103/PhysRevD.98.126001},
	eprint = {1801.07620},
	journal = {Phys. Rev.},
	number = {12},
	pages = {126001},
	primaryclass = {hep-th},
	slaccitation = {%%CITATION = ARXIV:1801.07620;%%},
	title = {{Circuit Complexity in Fermionic Field Theory}},
	volume = {D98},
	year = {2018}}

@article{Jefferson:2017sdb,
	archiveprefix = {arXiv},
	author = {Jefferson, Ro and Myers, Robert C.},
	doi = {10.1007/JHEP10(2017)107},
	eprint = {1707.08570},
	journal = {JHEP},
	pages = {107},
	primaryclass = {hep-th},
	slaccitation = {%%CITATION = ARXIV:1707.08570;%%},
	title = {{Circuit complexity in quantum field theory}},
	volume = {10},
	year = {2017}}

@article{Cai:2016xho,
	archiveprefix = {arXiv},
	author = {Cai, Rong-Gen and Ruan, Shan-Ming and Wang, Shao-Jiang and Yang, Run-Qiu and Peng, Rong-Hui},
	doi = {10.1007/JHEP09(2016)161},
	eprint = {1606.08307},
	journal = {JHEP},
	pages = {161},
	primaryclass = {gr-qc},
	slaccitation = {%%CITATION = ARXIV:1606.08307;%%},
	title = {{Action growth for AdS black holes}},
	volume = {09},
	year = {2016}}

@article{Brown:2015lvg,
	archiveprefix = {arXiv},
	author = {Brown, Adam R. and Roberts, Daniel A. and Susskind, Leonard and Swingle, Brian and Zhao, Ying},
	doi = {10.1103/PhysRevD.93.086006},
	eprint = {1512.04993},
	journal = {Phys. Rev.},
	number = {8},
	pages = {086006},
	primaryclass = {hep-th},
	slaccitation = {%%CITATION = ARXIV:1512.04993;%%},
	title = {{Complexity, action, and black holes}},
	volume = {D93},
	year = {2016}}

@article{Brown:2015bva,
	archiveprefix = {arXiv},
	author = {Brown, Adam R. and Roberts, Daniel A. and Susskind, Leonard and Swingle, Brian and Zhao, Ying},
	doi = {10.1103/PhysRevLett.116.191301},
	eprint = {1509.07876},
	journal = {Phys. Rev. Lett.},
	number = {19},
	pages = {191301},
	primaryclass = {hep-th},
	slaccitation = {%%CITATION = ARXIV:1509.07876;%%},
	title = {{Holographic Complexity Equals Bulk Action?}},
	volume = {116},
	year = {2016}}

@article{Stanford:2014jda,
	archiveprefix = {arXiv},
	author = {Stanford, Douglas and Susskind, Leonard},
	doi = {10.1103/PhysRevD.90.126007},
	eprint = {1406.2678},
	journal = {Phys. Rev.},
	number = {12},
	pages = {126007},
	primaryclass = {hep-th},
	slaccitation = {%%CITATION = ARXIV:1406.2678;%%},
	title = {{Complexity and Shock Wave Geometries}},
	volume = {D90},
	year = {2014}}
\bibliographystyle{JHEP}

\end{document}